# CORRELATED STRONTIUM AND BARIUM ISOTOPIC COMPOSITIONS OF ACID-CLEANED SINGLE MAINSTREAM SILICON CARBIDES FROM MURCHISON

NAN LIU[1,2,3], MICHAEL R. SAVINA[2,3], ROBERTO GALLINO[4], ANDREW M. DAVIS[1,2,5], SARA BISTERZO[4,6], FRANK GYNGARD[7], FRANZ KÄPPELER[8], SERGIO CRISTALLO[9,10], NICOLAS DAUPHAS[1,2,5], MICHAEL J. PELLIN[1,2,3,5] AND IRIS DILLMANN[11].

[1]Department of the Geophysical Sciences, The University of Chicago, Chicago, IL, 60637, USA;
lnsmile@uchicago.edu;
[2]Chicago Center for Cosmochemistry, Chicago, IL 60637, USA;
[3]Materials Science Division, Argonne National Laboratory, Argonne, IL 60439, USA;
[4]Dipartimento di Fisica, Università di Torino, 10125 Torino, Italy;
[5]Enrico Fermi Institute, The University of Chicago, Chicago, IL 60637, USA;
[6]INAF−Osservatorio Astrofisico di Torino, 10025 Pino Torinese, Italy;
[7]Laboratory for Space Sciences, Washington University, St. Louis, MO 63130, USA;
[8]Karlsruhe Institute of Technology, Institut für Kernphysik, Karlsruhe 76021, Germany;
[9]INAF, Osservatorio Astronomico di Collurania, 64100 Teramo, Italy;
[10]INFN-Sezione di Napoli, 80126 Napoli, Italy;
[11]TRIUMF, 4004 Westbrook Mall, Vancouver BC V6T 2A3, Canada.

## ABSTRACT

We present strontium, barium, carbon, and silicon isotopic compositions of 61 acid-cleaned presolar SiC grains from Murchison. Comparison with previous data shows that acid washing is highly effective in removing both strontium and barium contamination. For the first time, by using correlated $^{88}Sr/^{86}Sr$ and $^{138}Ba/^{136}Ba$ ratios in mainstream SiC grains, we are able to resolve the effect of $^{13}C$ concentration from that of $^{13}C$-pocket mass on *s*-process nucleosynthesis, which points towards the existence of large $^{13}C$-pockets with low $^{13}C$ concentrations in AGB stars. The presence of such large $^{13}C$-pockets with a variety of relatively low $^{13}C$ concentrations seems to require multiple mixing processes in parent AGB stars of mainstream SiC grains.

1. INTRODUCTION

Presolar grains were discovered in primitive meteorites over 25 years ago by their anomalous isotopic compositions, which are too widely variable to be explained only by known chemical and physical fractionation processes in the solar system (Bernatowicz et al. 1987; Zinner et al. 1987; Lewis et al. 1990). Instead, such anomalous isotopic compositions have proven to be signatures of nucleosynthesis in different types of stars, supporting the idea that these ~μm-size mineral grains formed in the winds of AGB stars and/or explosions of stars prior to formation of the solar system. These grains were injected to interstellar medium, were later incorporated in the solar system's parental molecular cloud, survived early destruction in the protosolar nebula, and were eventually incorporated into meteorites. Therefore, presolar grains provide invaluable information on stellar nucleosynthesis and Galactic Chemical Evolution (GCE) (Nittler et al. 1997; Zinner et al. 2006).

Among the wide variety of presolar minerals, SiC is the most extensively studied because: (1) it survives harsh acid dissolution of primitive meteorites, allowing relatively easy separation from bulk meteorites (Amari et al. 1994); (2) the acid-resistant property of SiC allows acid-cleaning of grains after separation in order to minimize potential contamination by solar system materials, with the aim of retrieving nucleosynthetic isotopic signatures directly inherited from parent stars (Liu et al. 2014a); (3) according to thermodynamic equilibrium calculations, SiC condenses only from carbon-rich gas (Lodders & Fegley 1995). It is therefore rare in solar system materials, allowing simple identification of presolar SiC. Based on extensive isotopic studies of both light ($A < 56$) and heavy ($A > 56$) elements, more than 90% of presolar SiCs are mainstream grains from low-mass Asymptotic Giant Branch (AGB) stars (Hoppe et al. 1994; Clayton & Nittler 2004; Davis 2011; Zinner 2014), which are the stellar site of the main *s*-process (slow neutron capture) nucleosynthesis (Gallino et al. 1990; Arlandini et al. 1999; Bisterzo et al. 2011). Thus, measurements of isotopic compositions in acid-cleaned mainstream SiC grains provide a unique opportunity to constrain AGB model calculations of the main *s*-process nucleosynthesis at precisions that greatly exceed the current capability of spectroscopic observations (Liu et al. 2014a). Unless noted otherwise, the *s*-process refers hereafter to the main *s*-process nucleosynthesis in low-mass AGB stars.

A good understanding of *s*-process nucleosynthesis is critical to understanding other nucleosynthetic processes for heavy elements. For instance, due to many difficulties in precisely predicting *r*-process (rapid neutron capture) nuclide abundances (e.g., Cowan et al. 1991; Goriely & Arnould 2001; Pearson et al. 2005), *s*-process abundances are commonly subtracted from the solar system abundances in order to obtain the *r*-process residual (Burbidge et al. 1957; Cameron 1957). In addition, more precise predictions for main *s*-process nuclide abundances can improve our understanding of the weak *s*-process in massive stars for nuclei up to the first *s*-process peak, whose predictions are quite uncertain due to their strong dependence on nuclear reaction rates that are uncertain in the relevant stellar temperature regime (Pignatari et al. 2010). Furthermore, another unknown component, the Light Element Primary Process (LEPP), may contribute to the solar abundances of nuclei in the mass region between strontium and barium (*e.g.*, Travaglio et al. 2004).

Detailed *s*-process calculations in AGB models, however, suffer from uncertainties in the two *s*-process neutron sources. Carbon-13, the major neutron source, burns radiatively at about 8 keV during interpulse periods, providing neutron densities ($\rho_n$) of $\sim 10^7 - 10^8$ cm$^{-3}$ on the timescale of ~20 ka *via* $^{13}$C($\alpha,n$)$^{16}$O; the minor neutron precursor $^{22}$Ne burns at about 23 keV during convective thermal pulses (TPs), providing a short neutron burst with a peak neutron density up to $\sim 10^9 - 10^{10}$ cm$^{-3}$ *via* the marginally activated $^{22}$Ne($\alpha,n$)$^{25}$Mg reaction (Gallino et al. 1998). The efficiency of the minor neutron source depends strongly on the $^{22}$Ne($\alpha,n$)$^{25}$Mg rate. This rate is not well estimated from nuclear physics experiments (Wiescher et al. 2012; Bisterzo et al. 2014), but can be constrained by measuring mainstream grains for abundances of nuclides affected by branch points along the *s*-process path (Lugaro et al. 2003; Liu et al. 2014a). These branching ratios depend strongly on the neutron density provided by the $^{22}$Ne($\alpha,n$)$^{25}$Mg reaction (Käppeler et al. 1990a). The process by which a $^{13}$C-pocket (a region enriched in $^{13}$C) can form in AGB stars is uncertain. It is unclear how hydrogen is mixed from the bottom of the convective envelope into the top of the helium intershell in order to form $^{13}$C *via* $^{12}$C($p,\gamma$)$^{13}$N($\beta^+\nu$)$^{13}$C (Straniero et al. 1995; Denissenkov & Tout 2003; Herwig et al. 1997; Cristallo et al. 2009). Thus, the $^{13}$C mass fraction (the concentration of $^{13}$C nuclei), and the size (the $^{13}$C-pocket mass) and the shape (the distribution of $^{13}$C nuclei) of the $^{13}$C-pocket cannot be derived without stellar evolution models having a level of detail beyond current computational capabilities and are frequently parameterized in *s*-process calculations. Nonetheless, those uncertain quantities can be constrained by measuring mainstream SiC grains for abundances of neutron-magic nuclides ($^{88}$Sr, $^{138}$Ba, and $^{208}$Pb in particular) (Liu et al. 2014a). In addition, information about the $^{13}$C-pocket can also be obtained by measuring the $^{92}$Zr/$^{94}$Zr ratio in mainstream SiC grains, because the $^{92}$Zr Maxwellian Averaged Cross Section (MACS) deviates from $1/v_T$ by > 30% from 8 to 23 keV, while the $^{94}$Zr MACS closely follows the $1/v_T$ rule (see details in Liu et al. 2014b).

In previous studies of single mainstream SiC grains using Torino postprocess AGB models, it was found that $^{13}$C-pockets with a flat $^{13}$C profile and pocket masses less than $5.3 \times 10^{-4}\ M_\odot$ were required to explain minor groups of mainstream SiC grains with unusual $^{138}$Ba/$^{136}$Ba and $^{92}$Zr/$^{94}$Zr isotope ratios (Liu et al. 2014a,b). Although the model predictions with these $^{13}$C-pockets can cover the whole range of mainstream grains for barium and zirconium isotope ratios, predictions with other types of $^{13}$C-pockets can still partially explain some of the grain data. Thus, we could not exclude the possibility that the $^{13}$C-pockets of parent AGB stars have a range of different $^{13}$C profiles and pocket masses.

As pointed out by Liu et al. (2014a), coupled measurements of $^{138}$Ba/$^{136}$Ba and $^{88}$Sr/$^{86}$Sr ratios in mainstream SiC grains are required to monitor the neutron flow at two bottlenecks along the s-process path, which allows investigation of the features of the $^{13}$C-pockets in AGB stars (Figure 2 of Käppeler et al. 2011). For this purpose, we simultaneously measured strontium and barium isotopic compositions in 82 acid-cleaned presolar SiC grains from Murchison using resonance ionization mass spectrometry (RIMS). One of the aims of this work was to minimize the solar system strontium and barium contamination in the Murchison meteorite. We succeeded in obtaining strontium and barium isotopic data in grains free of solar system contamination, which allows us to derive constraints on the $^{13}$C-pockets in low-mass AGB stars using the correlated $^{88}$Sr/$^{86}$Sr–$^{138}$Ba/$^{136}$Ba ratios in mainstream SiC grains.

2. EXPERIMENTAL METHODS AND TORINO POSTPROCESS AGB MODEL

*2.1. Isotope Measurements in Single Presolar SiC Grains*

Strontium and barium isotopes were measured with the CHARISMA instrument at Argonne National Laboratory (Savina et al. 2003). The standards for strontium and barium isotope measurements were SRM 855a (an aluminum casting alloy with 0.018 wt% strontium) and BaTiO$_3$, respectively. We ground the two standards together to form a well-mixed powder and then mixed it with conductive epoxy for mounting on an aluminum stub. Barzyk et al. (2007) developed a two-color resonant ionization scheme (RIS) for barium that is used in this study. For strontium, we modified the two-color RIS adopted by Nicolussi et al. (1998). A first photon (with $\lambda_1 = 460.86$ nm in vacuum) excites the transition $5s^2(^1S_0) \rightarrow 5s5p(^1P_1^0)$, while a second photon with $\lambda_2$ = 405.214 nm further excites the strontium atom to a strong autoionizing state above the ionization potential (IP) (Mende et al. 1995). Note that the wavelengths given in Nicolussi et al. (1998) are values in air. All transitions in both strontium and barium were well saturated with the beam intensities used in our measurements.

The presolar SiC grains in this study were acid-cleaned prior to mounting onto a piece of gold foil. This mount was previously used in the study of Liu et al. (2014a). A detailed description of the acid-cleaning procedure and the mounting method for presolar SiC grains is given in Levine et al. (2009) and Liu et al. (2014a). The isotopic compositions of strontium and barium in presolar SiC grains from Murchison were measured simultaneously. Eighteen of the 82 grains analyzed had so little strontium and barium that the desorbing laser fluence required to produce a signal was damaging the gold mount and resulted in significant backgrounds due to secondary ions. Therefore, no strontium or barium isotope data is available for these 18 grains.

After RIMS analysis, carbon and silicon isotopic compositions of the 64 grains were determined with the Cameca NanoSIMS 50 at Washington University, St. Louis, by rastering a primary $Cs^+$ beam over each grain and simultaneously collecting secondary ions of $^{12}C^-$, $^{13}C^-$, $^{28}Si^-$, $^{29}Si^-$, and $^{30}Si^-$. Two of the 64 grains were determined to be of type X and another to be of type Y (Hoppe et al. 1994, 1997). Of the remaining 61 grains, 50 grains were determined to be mainstream. The remaining 11 grains were completely consumed during the RIMS measurement and could not be classified by NanoSIMS; these are grouped as mainstream grains for purposes of discussion, since > 90% of SiC grains are mainstream (Hoppe et al. 1994; Nittler 2003).

For NanoSIMS measurements of carbon and silicon isotopes, terrestrial SiC aggregates were used as standards and measured in between every ten grain measurements in order to monitor and correct for instrumental drift. The uncertainties of the NanoSIMS data are calculated by including uncertainties from both counting statistics and the overall scatter on the measured standards. Uncertainties for the RIMS data are given by Equation (1) of Liu et al. (2014a), considering both counting (Poisson) and non-counting statistical uncertainties. The minor isotope $^{84}$Sr (0.7% in terrestrial samples) is not produced during *s*-process nucleosynthesis, whereas $^{86-88}$Sr are significantly overproduced, making its measurement in grains challenging. The $^{84}$Sr counts obtained from single mainstream SiC grains are extremely low and sensitive to background noise. We subtracted the average counts at masses 83 and 85 (no peak present) from the counts collected at mass 84 to obtain the corrected $^{84}$Sr counts. Thirteen grains had background corrections higher than that of the $^{84}$Sr signal; their δ($^{84}$Sr/$^{86}$Sr) values are given as upper limits without background subtraction. In addition, the δ($^{84}$Sr/$^{86}$Sr) values for six other grains are not reported due to the extremely low signal-to-noise ratios.

Carbon, silicon, strontium, and barium isotope ratios of the 61 mainstream and unclassified grains, if available, are reported in Table 1. Carbon isotope data are reported as isotope ratios; silicon, strontium, and barium isotope data are reported as δ-values, defined as the deviation in parts per thousand from isotope ratios measured in grains relative to the terrestrial isotope ratios measured in the standard (*e.g.*, δ($^{88}$Sr/$^{86}$Sr) = [($^{88}$Sr/$^{86}$Sr)$_{grain}$/($^{88}$Sr/$^{86}$Sr)$_{standard}$−1] ×1000). Carbon-12, $^{28}$Si, $^{86}$Sr, and $^{136}$Ba were chosen as the reference isotopes for carbon, silicon, strontium, and barium, respectively. All isotope ratios are reported with 2σ uncertainties in Table 1.

## 2.2. Torino Postprocess AGB Models

An in-depth description of the Torino postprocess AGB model is given by Gallino et al. (1998). Updates are reported by Bisterzo et al. (2014), including solar abundances from Lodders et al. (2009), and reaction rates from KADoNiS v0.3[1] and additional more recent publications. It has been widely recognized that a range of *s*-process strengths is required to explain the *s*-process scatter observed in peculiar *s*-rich stars at a given metallicity (*e.g.*, Busso et al. 2001, Abia et al. 2002) and the heavy-element isotopic compositions in presolar SiC grains (*e.g.*, Lugaro et al. 2003, Barzyk et al. 2007, Liu et al. 2014a,b). However, the physical reasons of this scatter are not clear yet. **This could be caused by variations in both the pocket parameters and initial stellar mass. Because varying the initial stellar mass in Torino postprocess AGB models cannot explain the range of isotope ratios observed in the grains from this study, we investigate the possibility that, for a single initial stellar mass, the observed range is caused by the simultaneous occurrence of different physical processes** (*e.g.*, overshooting, rotation, magnetic fields, gravity waves). Due to uncertainties in the formation process of the $^{13}$C-pocket (Section 1), the Torino postprocess AGB model treats the $^{13}$C mass fraction, the distribution of $^{13}$C within the $^{13}$C-pocket, and the $^{13}$C-pocket mass as variables in (see Liu et al. 2014a,b). It is hoped that combinations of these parameters that successfully explain isotopic compositions of presolar grains and elemental abundances in stars will motivate future AGB models capable of more realistic explanation of $^{13}$C-pocket formation. Motivated by the data, we did the tests below using the Torino postprocess models to obtain different *s*-process efficiencies for a fixed metallicity. A detailed description of the treatment of the $^{13}$C-pocket is given in Liu et al. (2014a). The parameters for the $^{13}$C-pocket adopted in Torino postprocess AGB models are listed in Table 2.

---

[1]KADoNiS v0.3: Karlsruhe Astrophysical Database of Nucleosynthesis in Stars, website http://www.kadonis.org/, version v0.3.

*2.2.1. Variation of $^{13}$C mass fraction and $^{13}$C distribution within the $^{13}$C-pocket*

Rotation is a good candidate for causing variations in the mass fraction and distribution of $^{13}$C within the $^{13}$C pocket. After a TDU, due to a sharp drop of the angular momentum, the Goldreich-Schubert-Fricke (GSF) instability is triggered at the inner border of the slow rotating AGB envelope (close to the deepest point previously attained by the fully convective envelope and almost fixed soon after a TDU). When hydrogen-burning reignites, the GSF instability starts to move inward because of the shallower mean molecular weight gradient (μ gradient) and spreads $^{14}$N-rich material within the $^{13}$C-pocket. The efficiency of the GSF instability is determined by the initial rotation speed of an AGB star. Thus, a range of *s*-process efficiencies can be obtained in AGB stars with different initial rotation speeds (Piersanti et al. 2013).

In the Torino postprocess AGB models, this mixing process is mimicked by assuming different $^{13}$C mass fractions and profiles within the pocket. The $^{13}$C mass fraction within the $^{13}$C-pocket is referred to as a *case*, with the notation D12 to ST designating the mass fraction relative to a standard (ST) case. The $^{13}$C mass fractions in D12 to ST cases are those in the ST case (in which the solar abundance pattern is well reproduced) divided (D) by the corresponding factors. The distribution of $^{13}$C within the $^{13}$C-pocket is another parameter and we evaluate two possibilities: a decreasing-with-depth $^{13}$C profile (Three-zone models) and a flat $^{13}$C profile (Zone-II models).

For instance, the $^{13}$C masses in the D1.5 cases of Three-zone_p2 and Zone-II_p2 models are

$$M(^{13}C)_{Three-zone\_p2} = \frac{X(^{13}C)_I}{1.5} \times (2 \times M_I) + \frac{X(^{13}C)_{II}}{1.5} \times (2 \times M_{II}) + \frac{X(^{13}C)_{III}}{1.5} \times (2 \times M_{III}) \quad (1)$$

$$M(^{13}C)_{Zone-II\_p2} = \frac{X(^{13}C)_{II}}{1.5} \times (2 \times M_{II}) \quad (2), \text{respectively.}^{2}$$

---

[2] $M_I$= 4.0×10$^{-4}$ $M_\odot$, $M_{II}$=5.3×10$^{-4}$ $M_\odot$, $M_{III}$=7.5×10$^{-6}$ $M_\odot$, $X(^{13}C)_I$= 3.2×10$^{-3}$, $X(^{13}C)_{II}$= 6.8×10$^{-3}$ and $X(^{13}C)_{III}$= 1.6×10$^{-2}$.

In the equations above, $X(^{13}C)_i$ and $M(^{13}C)_i$ are the $^{13}C$ mass fraction and the $^{13}C$-pocket mass of zone-i, respectively, in the ST case of the Three-zone or Zone-II models.

### 2.2.2. Variation of $^{13}C$-pocket mass

We also consider variation in the $^{13}C$-pocket mass because, for instance, magnetic buoyancy (Busso et al. 2012) or overshoot plus Eddington-Sweet (ES) instability (*i.e.*, meridional circulation) might result in a larger $^{13}C$-pocket with respect to that adopted in current AGB stars (usually $1\times10^{-3}\,M_\odot$). We take rotation as an example to show possible variations in the $^{13}C$-pocket mass in rotating AGB stars. After the quench of a TP, the ES instability occurs in the helium intershell because of its nearly homogenous composition resulting from the thorough mixing during the previous TP. After a TDU, the receding envelope leaves a variable hydrogen profile that acts as a barrier for meridional circulations (due to the steep µ gradient). For large enough initial rotation velocities, an inward mixing of hydrogen can occur and, if the hydrogen mass fraction does not exceed a few $10^{-4}$, $^{13}C$ production is favored with respect to that of $^{14}N$. Thus, the ES instability can produce a "stretched" $^{13}C$-pocket. Piersanti et al. (2013) concluded that the ES instability is efficient in low metallicity AGB stars (for all initial rotation speeds) and in fast rotating metal-rich stars. However, they also pointed out the current limitation of their simplified approach to treat rotation in the models and the huge uncertainties affecting the efficiency of rotation-induced instabilities.

The variation of the overall $^{13}C$-pocket mass is mimicked in the Torino postprocess AGB model tests by assuming different pocket masses. Models with varying $^{13}C$-pocket masses are designated as Three-zone_d2.5 to Three-zone_p8 for Three-zone models and Zone-II to Zone-II_p16 for Zone-II models, in the normal case (Three-zone or Zone-II) divided (d) or multiplied (p) by the corresponding factors in their names. In the Three-zone $^{13}C$-pocket, the pocket is divided into three zones: Zone-I, -II, and -III, within each of which the $^{13}C$ and $^{14}N$ mass fractions are constant while decreasing from Zone-III to Zone-I with depth. The minimum pocket mass considered has negligible effects on strontium and barium isotopes and the maximum pocket mass corresponds to the entire helium-intershell after a TP.

Nitrogen-14, a neutron poison in the $^{13}C$-pocket, is formed by proton capture *via* $^{13}C(p,\gamma)^{14}N$ and consumes neutrons *via* $^{14}N(n,p)^{14}C$. We consider that only a small number of protons diffuse into the top layers of the $^{12}C$-rich helium intershell so that only a minor fraction of the $^{13}C$ can capture a proton to produce $^{14}N$ (Gallino et al. 1998 and references therein).

Nitrogen-14 production therefore depends linearly on the $^{13}$C abundance and the $^{14}$N/$^{13}$C ratio is fixed at ~1/30 for all three zones. In our postprocess calculations, we did not consider the possibility of a more external "$^{14}$N-pocket" production in the helium intershell. Indeed, if more protons than $^{12}$C nuclei are present, $^{14}$N is preferentially synthesized and is accompanied by a much lower $^{13}$C abundance (see, *e.g.*, Cristallo et al. 2009). In the absence of rotation, the $^{14}$N-pocket is engulfed by the next TP, and mixed with the $^{14}$N nuclei from the hydrogen-burning ashes. Then a fraction of the $^{14}$N nuclei is converted to $^{22}$Ne via $^{14}$N$(\alpha,\gamma)^{18}$F$(\beta^+)^{18}$O$(\alpha,\gamma)^{22}$Ne prior to the release of the short neutron burst via $^{22}$Ne$(\alpha,n)^{25}$Mg. The net result is a marginal addition to the final $^{22}$Ne production with no influence on the *s*-process (we verified this by performing some tests with an additional $^{14}$N-pocket). The presence of rotation instead may induce partial mixing of the $^{14}$N-pocket with the inner $^{13}$C-rich zone in the interpulse phase before the release of the major neutron source, thus reducing the efficiency of the *s*-process to a certain degree.

Finally, we point out that other physical mechanisms, which might also be at work in these layers, cannot be "a priori" excluded. For instance, gravity waves might shape the transition between the convective envelope and the underlying radiative layers during a TDU (Denissenkov & Tout 2003). In addition, magnetic buoyancy is a good candidate to form the $^{13}$C-pocket (Busso et al. 2012) and possibly to shape it during the long interpulse phase. In fact, embedded toroidal magnetic flux tubes sustained by dynamo effects may affect a large portion of the helium intershell, leading to extended $^{13}$C-pockets with relatively low $^{13}$C concentrations. Rotating stellar model calculations including magnetic fields are highly desirable because of the strong interconnection between rotation and magnetic fields.

*2.3. Neutron Capture Cross-Sections of Nuclides in the Krypton-Strontium Region*

The $^{86,87,88}$Sr MACSs recently reevaluated in the preliminary version of KADoNiS v1.0 (Karlsruhe Astrophysical Database of Nucleosynthesis in Stars version v1.0, Dillmann et al. 2014)[3] are used in this study. Table 3 compares the new MACS values to the previously recommended values by KADoNiS v0.3 (Dillmann et al. 2006) and Koehler et al. 2000 (Erratum 2001), which are also plotted in Figure 1 for comparison. While there are only minor changes at $kT = 30$ keV, the differences at higher and lower energies are large due to the use of different energy-dependencies. The evaluation of the $^{88}$Sr MACS in KADoNiS v0.3 is based on an activation measurement at $kT = 25$ keV by Käppeler et al. (1990b), which was used to normalize unpublished time-of-flight (TOF) data from Oak Ridge (R. Macklin, private communication to H. Beer, 1989, hereafter Mac89). Koehler et al. (2000) later reported new $^{88}$Sr MACS values based on an improved TOF measurement with a wider incident neutron energy range, better resolution, and the use of metallic samples. In KADoNiS v1.0, a weighted average at $kT = 30$ keV of the values from Koehler et al. (2000), Käppeler et al. (1990b), Mac89, and de L. Musgrove et al. (1978) is used; the energy-dependence of the $^{88}$Sr MACS is based on the Koehler et al. (2000) values normalized by a factor of 1.025. As a result, the KADoNiS v1.0 energy-dependence is different compared to the previous one in KADoNiS v0.3, ranging from $-13\%$ at $kT = 5$ keV to $+18\%$ at $kT = 100$ keV (Table 3). Note that the $^{88}$Sr MACS from Koehler et al. (2000) was used in previous studies of Torino postprocess AGB models (*e.g.*, Lugaro et al. 2003; Käppeler et al. 2011; Bisterzo et al. 2011).

---

[3] Available online at http://exp-astro.physik.uni-frankfurt.de/kadonis1.0/index.php.

For the $^{86}$Sr and $^{87}$Sr MACSs, the two sets of recommended values by KADoNiS v0.3 and KADoNiS v1.0 are quite similar at $kT = 30$ keV (Table 3). The $^{86}$Sr MACS evaluation in KADoNiS v0.3 was based on the TOF measurements by Mac89. In KADoNiS v1.0, the new $^{86}$Sr MACS value is the average from recent evaluated libraries ENDF/B-VII.1 and JENDL-4.0 (Chadwick et al. 2011; Shibata et al. 2011). The evaluated energy-dependencies are different compared to the previously recommended value, which leads to discrepancies of up to 40% at $kT = 100$ keV, and an additional uncertainty is introduced from the deviation between the two libraries. This results in a higher uncertainty compared to the previously recommended value in KADoNiS v0.3, but the KADoNiS v1.0 value includes possible uncertainties due to the extrapolation to higher and lower energies. For the new $^{87}$Sr MACS, the weighted average of the TOF results from Mac89, Walter & Beer (1985), and Bauer et al. (1991) was used in KADoNiS v1.0, the same as in KADoNiS v0.3. These data were not considered in the new ENDF/B-VII.1 or JENDL-4.0 evaluation by Chadwik et al. (2011) and Shibata et al. (2011), but their energy-dependence was used after normalization by a factor of 1.15. This leads to a difference in the MACS at $kT = 5$ keV of $-17\%$ and at $kT = 100$ keV of $+13\%$, compared to the previous values in KADoNiS v0.3.

For the $^{85}$Kr MACS, only theoretical estimates existed until recently, which disagree with each other by a factor of two, or even by a factor of four to five for the evaluated libraries by using their own statistical model calculations (Rauscher & Thielemann 2000; Goriely 2005). The previously recommended value in KADoNiS v0.3 was therefore a "semi-empirical" estimate based on the measured cross sections in the neighborhood, with a relatively large estimated uncertainty (55±45 mb at 30 keV). The newly recommended $^{85}$Kr MACS in KADoNiS v1.0 at 30 keV is deduced from the inverse reaction $^{86}$Kr$(\gamma,n)^{85}$Kr by Raut et al. (2013), with the help of the TALYS code. This value is consistent with the KADoNiS v0.3 value within the large uncertainties, but the recommended value in KADoNiS v1.0 is about 35% higher.

## 3. RESULTS

### 3.1. s-Process Nucleosynthesis of Strontium and Barium Isotopes

#### 3.1.1. Strontium isotopes

In the krypton-rubidium-strontium region, two branch points lie at $^{85}$Kr and $^{86}$Rb (Käppeler et al. 1990a). Figure 2 shows the two principal *s*-process channels in this region (Nicolussi et al. 1998): (1) $^{85}$Kr$(\beta^-)^{85}$Rb$(n,\gamma)^{86}$Rb$(\beta^-)^{86}$Sr$(n,\gamma)^{87}$Sr$(n,\gamma)^{88}$Sr and (2) $^{85}$Kr$(n,\gamma)^{86}$Kr$(n,\gamma)^{87}$Kr$(\beta^-)^{87}$Rb$(n,\gamma)^{88}$Rb$(\beta^-)^{88}$Sr. At branch points, some $^{85}$Kr and $^{86}$Rb nuclei undergo $\beta^-$ decay ($^{85}$Kr$_{\beta-}$ & $^{86}$Rb$_{\beta-}$) while others capture neutrons ($^{85}$Kr$_n$ & $^{86}$Rb$_n$). The $^{85}$Kr$_n$/$^{85}$Kr$_{\beta-}$ and $^{86}$Rb$_n$/$^{86}$Rb$_{\beta-}$ ratios are therefore positively correlated with the peak neutron density provided by the $^{22}$Ne neutron source during advanced TPs and consequently with the $^{22}$Ne$(\alpha,n)^{25}$Mg reaction rate. In addition, the isomeric state $^{85}$Kr has a considerably shorter half-life ($^{85}$Kr$_{iso}$, $t_{1/2}$ = 4.5 hr) against $\beta^-$ decay than the ground state $^{85}$Kr ($^{85}$Kr$_g$, $t_{1/2}$ = 11 a). This isomer is not thermalized at the helium-shell temperature in AGB stars, so the $^{85}$Kr$_g$ and $^{85}$Kr$_{iso}$ are treated as independent nuclei (Ward et al. 1976; Walter et al. 1986). The production ratio of $^{85}$Kr$_{iso}$ relative to $^{85}$Kr$_g$ *via* $^{84}$Kr$(n,\gamma)^{85}$Kr is unity in low-mass AGB stars (Beer 1991). Due to its extremely short half-life, 80% of $^{85}$Kr$_{iso}$ undergoes $\beta^-$ decay to $^{85}$Rb, while the remaining 20% decays to $^{85}$Kr$_g$, which has a half-life of 11.5 a (Käppeler et al. 1990a).

Strontium has four isotopes: $^{84}$Sr, $^{86}$Sr, $^{87}$Sr, and $^{88}$Sr (Figure 2). The proton-rich nuclide $^{84}$Sr is off the *s*-process path and is therefore destroyed by neutron capture during the *s*-process in AGB stars. However, $^{84}$Sr is not destroyed in the envelopes of AGB stars and the low $^{84}$Sr/$^{86}$Sr ratios of mainstream grains are due to $^{86}$Sr overproduction, not $^{84}$Sr destruction. In addition to the strontium production by the weak *s*-process in massive stars (Pignatari et al. 2010), $^{86}$Sr, $^{87}$Sr, and $^{88}$Sr are all prevailingly made in AGB stars during the main *s*-process: 60% of solar system $^{86}$Sr, and $^{87}$Sr, and 70% of neutron-magic $^{88}$Sr (Travaglio et al. 2004; Bisterzo et al. 2014). The solar system $^{87}$Sr abundance receives additional radiogenic contribution from $^{87}$Rb $\beta^-$ decay ($t_{1/2}$ = 49.2 Ga), although this effect is minor in regard to the variation discussed here: the $^{87}$Sr/$^{86}$Sr ratio of the bulk solar system has increased only from 0.698 to 0.756 ($\delta^{87}$Sr by ~80‰) in the last 4.56 Ga.

The abundances of $^{86}$Sr, $^{87}$Sr, and $^{88}$Sr are affected by the branch points starting at $^{85}$Kr. The final $^{86}$Sr and $^{87}$Sr abundances are determined by the $^{22}$Ne neutron source, because the $^{86}$Sr and $^{87}$Sr produced in the $^{13}$C-pocket are partially destroyed during the peak neutron density induced by marginal activation of the $^{22}$Ne neutron source. The MACS of $^{88}$Sr is a factor of ten smaller than that of $^{86}$Sr because it is neutron-magic ($N = 50$). It lies at the reunification of the two channels and behaves as a bottleneck at the first s-process peak. Thus, its abundance is mainly affected by uncertainties in the major neutron source $^{13}$C.

### 3.1.2. Barium isotopes

A partial nuclide chart for the xenon-cesium-barium region is given in Figure 1 of Liu et al. (2014a) with the major branch point at $^{134}$Cs ($t_{1/2}$ =2.1 a). There are two principal s-process channels in this region in low-mass AGB stars: (1) $^{134}$Cs$(\beta^-)^{134}$Ba$(n,\gamma)^{135}$Ba$(n,\gamma)^{136}$Ba and (2) $^{134}$Cs$(n,\gamma)^{135}$Cs$(n,\gamma)^{136}$Cs$(\beta^-)^{136}$Ba. Therefore, the ratio of $^{134}$Cs$_n$/$^{134}$Cs$_{\beta-}$ is positively correlated with the $^{22}$Ne$(\alpha,n)^{25}$Mg rate. The branching ratio at $^{134}$Cs is temperature-sensitive, because the $^{134}$Cs $\beta^-$ decay rate is strongly temperature-dependent (see Section 4.4 of Liu et al. 2014a for details). Once $^{135}$Cs ($t_{1/2}$ =2.3 Ma) is formed by neutron-capture on $^{134}$Cs, it is stable on the timescale of the s-process occurring in the helium intershell (~20 ka).

Barium has seven isotopes: $^{130}$Ba and $^{132}$Ba are p-only, $^{134}$Ba and $^{136}$Ba s-only, $^{138}$Ba s-mostly, and $^{135}$Ba and $^{137}$Ba are s,r-mixed nuclides (Arlandini et al. 1999; Bisterzo et al. 2011, 2014). As shown in Figure 1 of Liu et al. (2014a), the $^{134}$Ba/$^{136}$Ba ratios in mainstream SiC grains lie above the solar system value because of the $^{134}$Cs branch point (and the fact that cesium did not condense quantitatively, see discussion in Section 4.6 of Liu et al. 2014a), although both nuclei are s-only isotopes. The two s-process paths join at $^{136}$Ba. Cesium-136 ($t_{1/2}$ = 13 d, independent of temperature), almost fully decays to $^{136}$Ba, and therefore predictions for the $^{136}$Ba and $^{137}$Ba abundances are unaffected by the branching effect. Similar to $^{88}$Sr, neutron-magic $^{138}$Ba behaves as a bottleneck at the second s-process peak because of its extremely small MACS. Its abundance is therefore mainly affected by uncertainties in the major neutron source $^{13}$C.

*3.2. Solar Contamination of Strontium and Barium in Presolar SiC Grains*

There is clear evidence for heavy-element contamination from solar system materials on presolar grain surfaces caused by laboratory procedures and/or secondary processes that occurred on the Murchison parent body (Barzyk et al. 2007; Marhas et al. 2007). If a presolar SiC grain is contaminated, the measured isotope ratios will be the result of mixing between the solar system isotope ratios and those of the parent stars, which prevents us from obtaining stringent constrains on stellar nucleosynthesis model calculations. Solar system barium contamination in presolar SiC grains was found in previous studies (Barzyk et al. 2007; Marhas et al. 2007). Solar strontium contamination in presolar SiC grains is also evident in the data of Nicolussi et al. (1998). Both strontium and barium belong to Group II alkaline elements and have similar chemical behavior.

We compare the barium and strontium isotopic compositions of 11 unclassified and 50 mainstream grains from this study in Figure 3. Two distinct groups of mainstream grains can be observed. Six mainstream grains (open squares, highlighted in Table 1) show close-to-solar barium and/or strontium isotopic compositions in the three-isotope plots in Figure 3. Five of the six grains (G32, G64, G214, G219, and G422) have strontium and barium isotope ratios that are close-to-solar within 2σ uncertainties, in contrast to the other 44 mainstream grains with extremely negative δ($^{84}$Sr/$^{86}$Sr) and/or δ($^{135}$Ba/$^{136}$Ba) values. The other grain (G189) did not have measureable barium. If these near-normal isotope ratios were caused by surface contamination, we would expect variation in both the isotope ratios and the elemental ratios from solar to *s*-process values at different depths in a single grain. No such trend was observed in any of the six grains analyzed by RIMS and the consistent close-to-solar δ($^{84}$Sr/$^{86}$Sr) and/or δ($^{135}$Ba/$^{136}$Ba) values in these six mainstream grains are therefore unlikely caused by surface contamination. Instead, it is highly likely that they were formed in parent AGB stars with extremely low mean neutron exposures (see discussion in Section 4.4). In addition, the 11 unclassified grains and the other 44 mainstream grains (black dots) in Figure 3 define a tight cluster and span the same range in both three-isotope plots. We therefore group the unclassified grains as mainstream for the following discussion.

We compare the acid-cleaned mainstream SiC grains from this study with previous literature data for barium and strontium isotopic compositions in Figure 4. For barium isotopes, our new data are in good agreement with the acid-washed SiC grain data from Liu et al. (2014a). While the two grains with close-to-solar barium isotopic compositions from Liu et al. (2014a) are unclassified, all five grains from this study lying in this region are classified as mainstream. For strontium isotopes, we compare our new grain data with the unwashed grain data from Nicolussi et al. (1998). These authors did not measure carbon or silicon isotopes, so these grains are unclassified. It is likely that most of the Nicolussi et al. (1998) grains are mainstream, because > 90% of presolar SiC grains are mainstream (Hoppe et al. 1994; Nittler 2003). The grains from Nicolussi et al. (1998) show a range of $\delta(^{84}Sr/^{86}Sr)$ values from above zero to slightly below −800‰ in Figure 4. Eight of the 26 grains (30%) measured by Nicolussi et al. (1998) have solar isotopic compositions. In contrast, only six grains in our study (10%) show close-to-solar $\delta(^{84}Sr/^{86}Sr)$ values and the remaining 48 mainstream and unclassified SiC grains all have $\delta(^{84}Sr/^{86}Sr)$ values less than −500‰ with the majority lower than −800‰. The difference in the distribution of $\delta(^{84}Sr/^{86}Sr)$ values between the two data sets is most likely caused by surface strontium contamination with solar system materials in the grains studied by Nicolussi et al. (1998).

### 3.3. $\delta(^{87}Sr/^{86}Sr)$ in Mainstream SiC Grains and Its Implication

As discussed in Section 3.1, $^{86}Sr$ and $^{87}Sr$ are along the same *s*-process path (Path 1 in Figure 2), and AGB model predictions for $\delta(^{87}Sr/^{86}Sr)$ barely deviate from the solar system value (Figure 5). The radioactive nuclide $^{87}Rb$, however, decays to $^{87}Sr$ with a half-life of 49.2 Ga, which could enhance $\delta(^{87}Sr/^{86}Sr)$ values in presolar SiC grains after grain formation in the outflow of AGB stars. Rubidium, however, is unlikely to condense into SiC due to its volatility. Liu et al. (2014a) showed that cesium did not condense into presolar SiC grains, and the volatilities of cesium and rubidium are comparable. Furthermore, given the long half-life of $^{87}Rb$, the amount of rubidium required in order to have a significant effect is large. The present-day $\delta(^{87}Sr/^{86}Sr)$ value of a grain depends on its initial $^{87}Rb/^{87}Sr$ ratio when it was formed, and the grain age $t$,

$$\delta(^{87}Sr/^{86}Sr)_{grain} = \left(\frac{^{87}Rb}{^{87}Sr}\right)_{initial} (1 - e^{\lambda t}) \times 1000 \quad (3),$$

where $\lambda$ is the decay constant. Thus, even if the grain is as old as 14 Ga, the initial $^{87}Rb/^{87}Sr$ ratio in the grain has to be enriched by a factor of about 80 relative to the ratio in the AGB stellar envelope (~0.2) in order to produce a $\delta(^{87}Sr/^{86}Sr)$ ~500‰. In fact, the volatility of rubidium is lower than that of strontium and the initial $^{87}Rb/^{87}Sr$ ratio therefore should be lower than 0.2 (*i.e.,* the Rb/Sr enrichment factor should be less than one). Therefore, the effect of $^{87}Rb$ decay to $^{87}Sr$ in presolar SiCs is negligible. Although the two grains (G222 and G334) show high $\delta(^{87}Sr/^{86}Sr)$ values (>400‰) in Figure 5, both have large statistical uncertainties due to extremely low strontium concentrations. The $\delta(^{87}Sr/^{86}Sr)$ values of G222 is within 2σ of the solar system value, while that of G334 is within 2.5σ.

### 3.4. Comparison between Single SiC Grains and SiC Aggregates

Strontium and barium isotopic compositions of presolar SiC grains have also been previously measured in SiC aggregates, which consist of millions of single presolar SiC grains, using Thermal Ionization Mass Spectrometry (TIMS) and Secondary Ion Mass Spectrometry (SIMS) (Ott & Begemann 1990a, b; Zinner et al. 1991; Prombo et al. 1993; Jennings et al. 2002; Podosek et al. 2004). In Figure 6, the best-fit lines with 95% confidence bands for both SiC aggregates (dashed line) and single SiC grains (solid line) are shown, which are calculated using the weighted Orthogonal Distance Regression (ODR) method[4]. Note that uncertainties in the aggregate fittings are insignificant so that they are not shown in the plots. To better compare with isotopic compositions of SiC aggregates that consist of different types of presolar grains, all the single grain data, including 61 mainstream, one Y and two X grains from this study are used for calculating the best-fit line for strontium; for barium, the grain data from Liu et al. (2014a) are also included. We also calculated the Mean Square Weighted Deviation (MSWD) values for the barium (0.57) and strontium (1.46) best-fit lines of our single grain data using the Isoplot software (Ludwig 2012). The close-to-unity MSWD values demonstrate that the deviation of grain data from the best-fit lines in the two plots is mainly caused by analytical uncertainties of the single grain data. Consistent with the conclusion for barium isotopes by Liu et al. (2014a), the best-fit lines for the single grains and aggregates agree well with each other in both three-isotope plots. This is also consistent with AGB model predictions that variations in the $^{137}Ba/^{136}Ba$ and $^{87}Sr/^{86}Sr$ production ratios in low-mass AGB stars should be limited (see discussion in Section 3.1).

Such linear correlations are not expected in the three-isotope plots of $\delta(^{88}Sr/^{86}Sr)$ versus $\delta(^{84}Sr/^{86}Sr)$, or $\delta(^{134}Ba/^{136}Ba)$ and $\delta(^{138}Ba/^{136}Ba)$ versus $\delta(^{135}Ba/^{136}Ba)$. This stems from the fact that one cannot define single "G-components (pure *s*-process isotopes made in the helium intershell)" for $\delta(^{134}Ba/^{136}Ba)$ and $\delta(^{88}Sr/^{86}Sr)$, and $\delta(^{138}Ba/^{136}Ba)$ due to the $^{134}Cs$ and $^{85}Kr$ branch points, and possibly, varying $^{13}$C-pockets in parent AGB stars. A good example is shown in Figure 1 of Prombo et al. (1993), in which the aggregate data are quite scattered in the plots of δ($^{134}$Ba/$^{136}$Ba) and δ($^{135}$Ba/$^{136}$Ba) versus δ($^{135}$Ba/$^{136}$Ba), in contrast to the well-defined linear fit to the data in the δ($^{137}$Ba/$^{136}$Ba) versus δ($^{135}$Ba/$^{136}$Ba) plot.

---

[4] The weighted ODR method is a linear least-squares fitting that minimizes scatter orthogonal to the best fit line and considers uncertainties in both x-axis and y-axis for each data point.

## 4. DISCUSSION

### *4.1. Constraints on $^{13}$C-Pockets in Low-Mass AGB Stars*

We first discuss the effects of initial stellar mass and metallicity on AGB model predictions for δ($^{88}$Sr/$^{86}$Sr). We use silicon isotope ratios as proxies to determine the effect of metallicity on *s*-process isotopic signatures in mainstream SiC grains in Section 4.1.2. As no clear correlation can be found between metallicity and *s*-process isotope ratios, the variation in the isotope ratios is more likely to be a result of varying $^{13}$C-pockets in parent AGB stars. Thus, we derive constraints on the $^{13}$C-pocket based on correlated δ($^{88}$Sr/$^{86}$Sr) and δ($^{138}$Ba/$^{136}$Ba) in Sections 4.1.3. The six mainstream SiC grains (open squares) with close-to-solar strontium and/or barium isotopic compositions are not included in Sections 4.1−4.3. The neutron density environments of their parent AGB stars will be discussed separately in Section 4.4.

### *4.1.1. Effect of AGB initial stellar mass and metallicity*

Previous isotopic studies of mainstream SiC grains constrain the masses of the parent stars to 1.5−3 $M_\odot$ and the metallicities to close-to-solar (Lugaro et al. 2003; Barzyk et al. 2007). We therefore compare our strontium data with Torino postprocess AGB model predictions for AGB stars with masses and metallicities within this range. The $^{22}$Ne(α,n)$^{25}$Mg reaction rate in the model is the lower limit of the recommended value from Käppeler et al. (1994, K94 rate hereafter). Unless noted otherwise, the K94 rate is adopted in model calculations.

As shown in Figure 7, the dependence of model predictions for δ($^{88}$Sr/$^{86}$Sr) on the initial stellar mass results from the fact that the $^{86}$Sr destruction in AGB stars depends slightly on the $^{22}$Ne(α,n)$^{25}$Mg rate, because the $^{86}$Sr nuclei produced in the $^{13}$C-pocket are partially destroyed during the peak neutron density induced by the marginal activation of the $^{22}$Ne(α,n)$^{25}$Mg reaction at the bottom of the helium-burning zone during a TP. The peak temperature increases with increasing core mass (Straniero et al. 2003), resulting in more efficient operation of the $^{22}$Ne(α,n)$^{25}$Mg reaction in 3 $M_\odot$ AGB stars. This leads to relatively higher $^{86}$Sr destruction in the 3 $M_\odot$, $Z_\odot$ case and therefore ~150‰ higher δ($^{88}$Sr/$^{86}$Sr) predictions compared to the 2 $M_\odot$, $Z_\odot$ case; the variation in $\delta(^{84}Sr/^{86}Sr)$ values is negligible due to the large overproduction of $^{86}Sr$ in the *s*-process.

Model predictions for $\delta(^{88}Sr/^{86}Sr)$ values depend strongly on the initial stellar metallicity, because the *s*-process efficiency depends on the number of $^{13}C$ nuclei per iron seed. As $^{13}C$ is primary, produced by proton capture onto the freshly synthesized $^{12}C$ nuclei in the helium intershell, the *s*-process efficiency depends linearly on the initial metallicity. For instance, Three-zone model predictions of 2 $M_\odot$, $Z_\odot$ AGB stars in the D1.5 case are comparable to those of 2 $M_\odot$, 0.5 $Z_\odot$ AGB stars in the D3 case because the number of iron seeds in $Z_\odot$ AGB stars is twice of that in 0.5 $Z_\odot$ AGB stars. One-half-$Z_\odot$ AGB models, however, predict a longer carbon-rich phase and ~50‰ higher $\delta(^{88}Sr/^{86}Sr)$ value for the last several TPs because: (1) the convective envelope of 0.5 $Z_\odot$ AGB stars starts with less oxygen and therefore becomes carbon-rich after fewer pulses than that of $Z_\odot$ AGB stars; and (2) the lower the metallicity, the higher the core mass and in turn, the higher the temperature attained during TPs. Thus, the higher $^{86}Sr$ depletion in the 0.5 $Z_\odot$ AGB stars results in its lower final abundance and therefore a slightly higher $\delta(^{88}Sr/^{86}Sr)$ value. To summarize, $\delta(^{88}Sr/^{86}Sr)$ predictions are strongly affected by the initial stellar metallicity.

The effect of the initial stellar metallicity, however, can be compensated by varying the $^{13}C$ mass fraction (D3 to U2 cases) in Torino postprocess AGB models. In the following Sections, we use 2 $M_\odot$, 0.5 $Z_\odot$ AGB models as representative to derive constraints on the $^{13}C$-pocket mainly because of its longer carbon-rich phase. Thus, the potential grain condensation regime is more extended, which can lead to a better overlap of model predictions with the grain data. **Finally, we point out that the mass loss rate is quite uncertain in AGB stellar models. The more efficient mass loss rate reported in recent studies[5] will be incorporated into future Torino postprocess AGB models to investigate the effect on model predictions. At the moment, we cannot exclude the possibility that a stronger mass loss rate would have important consequences on the final *s*-process isotopic ratios of the convective envelope.**

*4.1.2. Mainstream grain data range: Variation in the $^{13}C$-pocket or initial metallicity?*

The *s*-process efficiency correlates positively with the number of $^{13}C$ nuclei within the $^{13}C$-pocket, which depends on $^{13}C$ profile, $^{13}C$-pocket mass, and $^{13}C$ mass fraction, and correlates inversely with the number of iron seeds, which in turn depends on initial stellar metallicity. Thus, for a fixed-mass $^{13}$C-pocket with a fixed $^{13}$C profile, the *s*-process nuclide production depends on both the $^{13}$C mass fraction and the initial stellar metallicity. In turn, the spread of mainstream grain data (*e.g.*, in Figure 7) could be caused by variation in the $^{13}$C-pocket (*e.g.*, a range of $^{13}$C-mass fractions) and/or in the initial stellar metallicity.

---

[5] This mass loss rate is used in more recent evolutionary AGB models (Cristallo et al. 2011) and is needed to reproduce the revised luminosity function of Galactic carbon stars (Guandalini & Cristallo 2013).

In this Section, we discuss the effects of the initial metallicity of parent AGB stars on model predictions of strontium and barium isotope compositions in mainstream SiC grains. Delta-($^{29,30}$Si/$^{28}$Si) values in mainstream SiC grains are proxies of initial metallicities of their parent AGB stars (Lugaro et al. 1999; Zinner et al. 2006). It was shown that the spread of mainstream grain data in the plot of δ($^{29}$Si/$^{28}$Si) versus δ($^{30}$Si/$^{28}$Si) does not correspond only to the spread of the *s*-process "G-component". Instead, the spread is dominated by the GCE effect for stellar metallicities around solar. In particular, $^{28}$Si is a primary isotope, mostly produced by SNII, for which the $^{28}$Si/Fe ratio in the thin Galactic disk decreases with increasing metallicity (similar to other alpha isotopes, like $^{16}$O, $^{24}$Mg, $^{40}$Ca, $^{48}$Ti, Kobayashi & Nakasato 2011). On the other hand, $^{29}$Si and $^{30}$Si are secondary isotopes whose nucleosynthesis depends on the abundances of primary isotopes and their abundances increase linearly with increasing metallicity. Thus, the higher the δ($^{29,30}$Si/$^{28}$Si) values, the higher the metallicity (Lugaro et al. 1999; Zinner et al. 2006). As shown in Figure 8a, δ($^{29,30}$Si/$^{28}$Si) values of the mainstream SiC grains from Liu et al. (2014a) and this study all plot along a line with the slope of 1.4, in agreement with mainstream SiC grain data from the WUSTL presolar database[6] (Hynes & Gyngard 2009).

We plot δ($^{88}$Sr/$^{86}$Sr) and δ($^{138}$Ba/$^{136}$Ba) against δ($^{29,30}$Si/$^{28}$Si) values in Figures 8b−e to examine if there exists any correlation. No simple linear correlation can be observed between δ($^{138}$Ba/$^{136}$Ba) and δ($^{29,30}$Si/$^{28}$Si) values. We tried weighted ODR fits for δ($^{88}$Sr/$^{86}$Sr) versus δ($^{29,30}$Si/$^{28}$Si) values to look for possible correlations. The slopes of the fitting lines are 0.38±0.34 and 0.19±0.45 with 95% confidence, which excludes the possibility of even weak correlations. Thus, even if the parent AGB stars had varying initial metallicities, this may not be the dominant factor in determining the range of strontium and barium isotopic compositions seen in mainstream SiC grains. Variation in metallicities of parent AGB stars, however, could help in enlarging the range spanned by AGB model calculations. Assuming parent AGB stars with a fixed metallicity (next Section) yields constraints on the $^{13}$C-pocket corresponding to the maximum variation allowed by the grain data.

---

[6] WUSTL presolar database, website http://presolar.wustl.edu/~pgd/.

### 4.1.3. Constraints on the $^{13}$C-pocket
#### 4.1.3.1. Smoking Guns: Correlated $\delta(^{88}Sr/^{86}Sr)$ and $\delta(^{138}Ba/^{136}Ba)$

In this Section, we derive constraints on the $^{13}$C-pocket by comparing the grain data from this study with AGB models with a fixed metallicity ($Z = 0.5\ Z_\odot$). We first summarize effects of different sources of uncertainties on AGB model predictions for $\delta(^{138}Ba/^{136}Ba)$, with respect to $\delta(^{88}Sr/^{86}Sr)$. Comparison of AGB model predictions with acid-cleaned mainstream SiC grains for barium isotopes is given in detail in Liu et al. (2014a) and therefore is not repeated here.

(1) The neutron-magic nuclei $^{88}$Sr ($N = 50$) and $^{138}$Ba ($N = 82$) sit at the first and second s-process peaks, respectively, along the s-process path. They behave as bottlenecks in the main s-process path, and regulate the neutron flow in the s-process path (Figure 2 of Käppeler et al. 2011). This is because both isotopes have extremely small MACSs at AGB stellar temperatures (e.g., $\sigma_{MACS}{}^{138}Ba = 4.00\pm0.20$ mb, $\sigma_{MACS}{}^{88}Sr = 6.13\pm0.11$ mb at 30 keV), which are a factor of ten lower than those of their normalization s-only isotopes, $^{136}$Ba and $^{86}$Sr, respectively. Consequently, first-order equilibrium cannot be established at these bottlenecks because of their much lower MACSs (Clayton 1968). Thus, $^{88}$Sr and $^{138}$Ba can accumulate relative to their reference isotopes depending on the details of the $^{13}$C-pocket adopted in AGB models (Liu et al. 2014a).

(2) AGB model predictions for $\delta(^{138}Ba/^{136}Ba)$ are unaffected by uncertainties in the $^{22}Ne(\alpha,n)^{25}Mg$ rate because different paths (due to the $^{134}$Cs branch point) rejoin at $^{136}$Ba in the cesium and barium mass region (see discussion in Section 3.1). In addition, uncertainties in the neutron capture MACSs of $^{136}$Ba and $^{138}$Ba are $\leq \pm5\%$, corresponding to $\pm50‰$ uncertainty for $\delta(^{138}Ba/^{136}Ba)$ model predictions at most. On the contrary, predictions for $\delta(^{88}Sr/^{86}Sr)$ suffer from the uncertain $^{22}Ne(\alpha,n)^{25}Mg$ rate. Although the $^{88}$Sr MACS is well measured with only $\pm3\%$ uncertainty from 5 to 10 keV, there exists $\pm100‰$ uncertainty in model predictions for $\delta(^{88}Sr/^{86}Sr)$ due to $\sim\pm10\%$ uncertainty for the $^{86}$Sr MACS (Table 3).

(3) AGB model predictions for both $\delta(^{88}Sr/^{86}Sr)$ and $\delta(^{138}Ba/^{136}Ba)$ are sensitive to the details of the $^{13}$C-pocket adopted in model calculations. As a general rule, Zone-II models predict a wider range of $\delta(^{88}Sr/^{86}Sr)$ and $\delta(^{138}Ba/^{136}Ba)$ values than Three-zone models (Liu et al. 2014b). In addition, model predictions for $\delta(^{138}Ba/^{136}Ba)$ decrease with decreasing $^{13}$C-pocket mass, while the dependence of $\delta(^{88}Sr/^{86}Sr)$ model predictions on the $^{13}$C-pocket mass are strong functions of both the $^{13}$C profile and the $^{13}$C mass fraction.

*4.1.3.2. Comparison of model predictions with grain data*

We compare $\delta(^{88}Sr/^{86}Sr)$ and $\delta(^{138}Ba/^{136}Ba)$ for 47 mainstream grains with Three-zone and Zone-II predictions with varying $^{13}$C-pocket masses in Figure 9. As the Zone-II $^{13}$C-pocket mass is about one half of the Three-zone one (Equations 1, 2), each vertical pair of plots in Figure 9 corresponds to approximately equal-mass $^{13}$C-pockets with different $^{13}$C profiles. As can be seen in Figure 9, AGB model predictions for both $\delta(^{88}Sr/^{86}Sr)$ and $\delta(^{138}Ba/^{136}Ba)$ depend strongly on the three parameters characterizing the $^{13}$C-pocket: the $^{13}$C-profile, the $^{13}$C-pocket mass, and the $^{13}$C mass fraction, $X(^{13}C) = M(^{13}C) / M(^{13}C\text{-pocket})$. A single $^{13}$C-pocket with both fixed pocket mass and $^{13}$C mass fraction apparently cannot explain the entire variation of the grain data in Figure 9. Also, as discussed in Section 4.1.2, metallicity is not the dominant factor in determining the isotope ratios in mainstream grains. Thus, it is highly likely that different physical processes in stellar interiors lead to formation of diverse $^{13}$C-pockets in parent AGB stars of mainstream SiC grains.

The $^{13}$C mass fraction is the most important parameter in determining the grain scatter in Figure 9. If the $^{13}$C mass fraction and pocket mass are fixed, the variation in isotopic compositions that results from changing the $^{13}$C profile is too limited to explain the scatter observed in grains; if the $^{13}$C profile and the $^{13}$C mass fraction are fixed, the best match with the grain data would be given by Three-zone $^{13}$C-pockets in the D2 case with the $^{13}$C-pocket mass varying from $1\times10^{-3}$ to $8\times10^{-3}$ $M_\odot$, which still fails to match many of the grains; if the $^{13}$C profile and the $^{13}$C-pocket mass are fixed, most of the $^{13}$C-pockets with a range of $^{13}$C mass fractions can well explain the grain scatter. Thus, for the three variables, the order of importance in explaining the variation of isotopic compositions in mainstream SiC grains is $^{13}$C mass fraction, $^{13}$C-pocket mass, and $^{13}$C profile within the $^{13}$C-pocket. We therefore plot the fixed-mass Three-zone or Zone-II $^{13}$C-pockets with varying $^{13}$C mass fractions in each panel of Figure 9.

Interesting patterns of model predictions are found in Figure 9, which depend strongly on the interplay among the three parameters of the $^{13}$C-pocket. AGB model predictions for $\delta(^{88}Sr/^{86}Sr)$ increase with increasing $^{13}$C mass fraction, while predictions for $\delta(^{138}Ba/^{136}Ba)$ show a highly nonlinear dependence on the $^{13}$C mass fraction, resulting in the coiled shapes. For Three-zone models, δ($^{88}$Sr/$^{86}$Sr) increases with increasing $^{13}$C-pocket mass in all cases, while the dependence of δ($^{138}$Ba/$^{136}$Ba) predictions on the $^{13}$C-pocket mass is also a function of the $^{13}$C mass fraction (Figures 9a−d). In contrast, variations in Zone-II model predictions for both δ($^{88}$Sr/$^{86}$Sr) and δ($^{138}$Ba/$^{136}$Ba) are more subtle, and are strong functions of both the $^{13}$C mass fraction and $^{13}$C-pocket mass. For instance, the Zone-II model predictions distinguish themselves from Three-zone in the cases above D2 (*e.g.*, D1.5 and D1.3): Zone-II predictions remain unchanged in the D1.5 case with increasing $^{13}$C-pocket mass until $2\times10^{-3}$ $M_{\odot}$ (Figures 9e−g). This difference results from the fact that by increasing the $^{13}$C mass fraction from D1.5 to D1.3, the increased amount of neutrons starts to flow more efficiently to the $^{138}$Ba bottleneck at the second *s*-process peak in Figures 9e−g, which leads to slightly lower δ($^{88}$Sr/$^{86}$Sr) and much higher δ($^{138}$Ba/$^{136}$Ba) predicted by the D1.3 case. Once the pocket mass is increased to $8\times10^{-3}$ $M_{\odot}$, the neutron exposure is more efficient and the *s*-flow reaches both bottlenecks, leading to both much higher δ($^{88}$Sr/$^{86}$Sr) and δ($^{138}$Ba/$^{136}$Ba) values in Figure 9h.

The isotopic compositions of mainstream SiC grains in this study define a tight cluster (grain-concentrated region) in Figure 9, with δ($^{88}$Sr/$^{86}$Sr) ranging from −200‰ to 0‰ and δ($^{138}$Ba/$^{136}$Ba) from −400‰ to −200‰. As shown in Figure 9, lower-than-D2 cases play important roles in matching the grain-concentrated region. Model predictions with even lower-mass $^{13}$C-pockets ($^{13}$C-pocket mass $<5\times10^{-4}$ $M_{\odot}$) are not shown in Figure 9, because predictions for strontium and barium isotope ratios barely change below this pocket mass except those for δ($^{84}$Sr/$^{86}$Sr), which continue to increase with decreasing pocket mass until $1.3\times10^{-4}$ $M_{\odot}$. On the other hand, the constraint on the maximum pocket mass comes from the mass of the helium intershell. As the $^{13}$C-pocket forms within the helium intershell, apparently the pocket mass has to lie below the helium-intershell mass, which decreases from $1\times10^{-2}$ to $8\times10^{-3}$ $M_{\odot}$ from the first to last TP during the AGB phase (Gallino et al. 1998; Straniero et al. 2003). As the pocket mass is constant for different TPs in Torino postprocess AGB models, the maximum $^{13}$C-pocket masses allowed in Torino postprocess AGB models are those in the Three-zone_p8 and Zone-II_p16 calculations (Figures 9d and 9h, respectively). In Figures 9d&h, Three-zone and Zone-II model predictions above the D4.5 case are shifted to both higher δ($^{88}$Sr/$^{86}$Sr) and δ($^{138}$Ba/$^{136}$Ba) values, while calculations below D4.5 show coiled shapes, avoiding the grain-concentrated region. Thus, only model calculations with lower-than-D4.5 mass fractions (Figures 9d&h) can reach the grain regime during carbon-rich phases. Constraints on the $^{13}$C mass fraction for $(1-2)\times10^{-3}$ $M_\odot$ $^{13}$C-pockets are summarized in Table 4.

AGB model predictions for $\delta(^{88}$Sr/$^{86}$Sr) suffer from uncertainties in the $^{22}$Ne($\alpha$,n)$^{25}$Mg rate, because predictions for the $^{86}$Sr abundance are affected by branch points in the krypton-rubidium-strontium region (see discussion in Section 3.1). In the next Section, we will evaluate the effects of uncertainties in both the $^{22}$Ne($\alpha$,n)$^{25}$Mg rate and the $^{86}$Sr MACS on our constraints in Table 4. Furthermore, AGB model predictions for $\delta(^{88}$Sr/$^{86}$Sr) are unaffected by adopting the Raut et al. (2013) $^{85}$Kr MACS in Torino postprocess AGB models, mainly because most of $^{85}$Kr undergoes $\beta^-$ decay instead of neutron-capture. We point out that although such nuclear uncertainties could systematically shift the constraints, two first-order conclusions are unaffected: (1) varying $^{13}$C-pockets with pocket mass $<1\times10^{-3}$ $M_\odot$ are insufficient to cover the whole range of the grain data and higher-mass $^{13}$C-pockets ($\geq1\times10^{-3}$ $M_\odot$) are required; and (2) in order to reach the grain data region, models must adopt inversely correlated pocket mass and $^{13}$C mass fraction.

*4.1.3.3. Effects of nuclear uncertainties*

The $^{22}$Ne($\alpha$,n)$^{25}$Mg rate is still quite uncertain. Although Longland et al. (2012) reported a small error (20%) for the $^{22}$Ne($\alpha$,n)$^{25}$Mg rate, their Monte Carlo statistical calculations were based on the old rate measurements by Jaeger et al. (2001). New direct experiments are needed to measure the resonances with higher accuracy. We therefore show the effects of the $^{22}$Ne($\alpha$,n)$^{25}$Mg rate by comparing the Three-zone calculations in Figures 9b, 10a&b, in which the $^{22}$Ne($\alpha$,n)$^{25}$Mg rate varies from K94 to ¼×K94. These rates are shown because the $^{22}$Ne($\alpha$,n)$^{25}$Mg rate is constrained to lie between ¼×K94 and K94 by $\delta(^{134}$Ba/$^{136}$Ba) values in mainstream SiC grains (Liu et al. 2014a). As shown in these figures, Three-zone predictions for $\delta(^{88}$Sr/$^{86}$Sr) for the last several TPs are decreased by ~200‰ and ~100‰, when the ¼×K94 and the ½×K94 rates are adopted, respectively. The final $^{88}$Sr abundance is determined by both the $^{13}$C and the $^{22}$Ne neutron sources, in proportion to the neutron exposures provided by each source. In contrast, the $^{86}$Sr (and $^{87}$Sr) final abundance is partially depleted during the activation of the $^{22}$Ne neutron source. Note that the recently recommended $^{22}$Ne($\alpha$,n)$^{25}$Mg rates by nuclear experiments lie between ½×K94 and K94 rates at AGB stellar temperatures (Jaeger et al. 2001; Longland et al. 2012), the lower limit ¼×K94 rate is inconsistent with these nuclear experiments and will not be considered in the following discussion.

AGB model predictions for $\delta(^{88}Sr/^{86}Sr)$ also depend on the $^{86}Sr$ MACS (±10% uncertainty, Table 3). Adopting 90% of the $^{86}Sr$ MACS recommended by KADoNiS v1.0 increases the final $^{86}Sr$ abundance, resulting in a systematic lowering of all model predictions for $\delta(^{88}Sr/^{86}Sr)$ by 100‰, as seen by comparing Figures 10b&d. In contrast, comparison of Figures 9b and 10c shows that adopting 110% of the KADoNiS v1.0 $^{86}Sr$ MACS increases the $\delta(^{88}Sr/^{86}Sr)$ predictions by varying degrees from D3 to D1.3 cases (≤100‰). For reference, predictions for $\delta(^{88}Sr/^{86}Sr)$ with 110% of the KADoNiS v1.0 $^{86}Sr$ MACS are increased by only 50‰ at the first carbon-rich TP and by 90‰ at the last TP in the D3 case. This results from the fact that for lower $^{13}C$ mass fraction cases, the final $^{86}Sr$ abundance in the convective envelope, from which the grains derived, receives less contribution from the "G-component" and therefore a higher relative contribution from the "N-component (initial isotopes present in the convective envelope)". Consequently, the increase in the $\delta(^{88}Sr/^{86}Sr)$ prediction for the convective envelope after each TP is less than 100‰ in the D3 and D2 cases in Figure 10c in comparison to Figure 9b.

Considering uncertainties in both the $^{22}Ne(\alpha,n)^{25}Mg$ rate and the $^{86}Sr$ MACS, the two extreme cases of the $\delta(^{88}Sr/^{86}Sr)$ prediction can be obtained by adopting the K94 rate and 110% KADoNiS v1.0 $^{86}Sr$ MACS (Figure 10c), and the ½×K94 rate and 90% KADoNiS v1.0 $^{86}Sr$ MACS (Figure 10d). Different sets of nuclear inputs slightly shift our constraints on the $^{13}C$-pocket. For instance, when the ½×K94 and 90% KADoNiS v1.0 $^{86}Sr$ MACS are adopted, Three-zone (Figure 10d) and Zone-II_p2 predictions agree poorly with the grain data (less than 30% of the grains are matched within uncertainties), and our first-order conclusion, that 88Sr accumulation depends on the details of the $^{13}C$-pocket, is therefore more strongly supported. Shifted constraints on the $^{13}C$ mass fraction for $(1-2)\times10^{-3}$ $M_\odot$ $^{13}C$-pockets are summarized in Table 5, for comparison with Table 4. As can be seen, our constraints on $(1-2)\times10^{-3}$ $M_\odot$ $^{13}C$-pockets in Table 4 are only slightly affected by the nuclear uncertainties.

*4.2. Further Constraints from the $^{135}Ba/^{136}Ba$ and $^{84}Sr/^{86}Sr$ ratios*

Figure 11 shows that Three-zone and Zone-II models with low total $^{13}C$ (low $^{13}C$-pocket mass and low $^{13}C$ mass fractions) predict $\delta(^{135}Ba/^{136}Ba)$ values too high to match the grains, while models with high total $^{13}C$ (high $^{13}C$-pocket mass and high $^{13}C$ mass fractions) predict $\delta(^{135}Ba/^{136}Ba)$ values too low to match the grains. Thus, further constraints on Table 4 from $\delta(^{135}Ba/^{136}Ba)$ values in the grains can be derived and are summarized in Table 6.

The $^{84}$Sr/$^{86}$Sr ratio represents the ratio of the dredged-up "G-component" to the "N-component" present in the convective envelope when the grains condensed (Lugaro et al. 2003). This is because the pure-*p* isotope $^{84}$Sr is completely destroyed by neutron capture in the helium intershell of AGB stars, while $^{86}$Sr is abundantly produced in the *s*-process. Figure 12 shows a subset of mainstream SiC that excludes those with very large uncertainties in δ($^{84}$Sr/$^{86}$Sr) because of the low abundance of $^{84}$Sr. The exclusion criteria are: (1) grains with less than 2000 measured strontium counts; (2) grains with only the δ($^{84}$Sr/$^{86}$Sr) upper limits determined due to low signal-to-noise ratios in their mass spectra. Within uncertainties, the grain data are in good agreement with the model predictions for carbon-rich envelopes. Thus, our derived constraints on the $^{13}$C-pockets in Table 6 are supported by δ($^{84}$Sr/$^{86}$Sr) values in mainstream SiC grains, which have total $^{13}$C masses of (6−19)×10$^{-7}$ $M_\odot$ for Three-zone models and (8−34)×10$^{-7}$ $M_\odot$ for Zone-II models.

*4.3. Constraints from This Study: Contradictions with Previous Conclusions?*

In Liu et al. (2014b), it was found that AGB model predictions for δ($^{92}$Zr/$^{94}$Zr) decrease with increasing $^{13}$C-pocket mass. It was concluded that a smaller Zone-II $^{13}$C-pocket is preferred in explaining the close-to-solar δ($^{92}$Zr/$^{94}$Zr) values found by Nicolussi et al. (1997) and Barzyk et al. (2007), in conjunction with δ($^{138}$Ba/$^{136}$Ba) below −400‰ in a few mainstream SiC grains found by Liu et al. (2014a). In fact, as pointed out in our previous studies, in most of the cases, it is impossible to separate the effect of the $^{13}$C mass fraction from that of the $^{13}$C-pocket mass by using only one tracer of the $^{13}$C-pocket (*e.g.*, δ($^{138}$Ba/$^{136}$Ba) or δ($^{92}$Zr/$^{94}$Zr)).

Previously, we considered only the $^{13}$C mass fractions from the D3 to U2 cases due to the fact that the *s*-process efficiency of the $^{13}$C-pocket with pocket mass below 1×10$^{-3}$ $M_\odot$ becomes too low to account for the range of δ($^{135}$Ba/$^{136}$Ba) and δ($^{96}$Zr/$^{94}$Zr) in mainstream SiC grains. As AGB model predictions for δ($^{88}$Sr/$^{86}$Sr) are sensitive to the $^{13}$C mass fraction, δ($^{88}$Sr/$^{86}$Sr) values in mainstream SiC grains from this study well constrain the $^{13}$C mass fraction to lie below D1.3. Once the $^{13}$C mass fraction is fixed to this lower value, we found that a larger $^{13}$C-pocket with lower $^{13}$C mass fraction than previously considered can also account for variations in δ($^{138}$Ba/$^{136}$Ba) and δ($^{92}$Zr/$^{94}$Zr) in mainstream SiC grains. For instance, in addition to Three-zone_d2.5 predictions in the D2 case found by Liu et al. (2014b), Three-zone_p2 predictions in the D6 case can also match the grains with δ($^{92}$Zr/$^{94}$Zr) > −50‰. The δ($^{88}$Sr/$^{86}$Sr) values in the mainstream SiC grains clearly exclude small $^{13}$C-pockets with concentrated $^{13}$C nuclei as the dominant pocket type and favor large $^{13}$C-pockets with diluted $^{13}$C nuclei in parent AGB stars.

Previously, Liu et al. (2014a) found that $\delta(^{138}\text{Ba}/^{136}\text{Ba})$ below −400‰ in a few mainstream grains can be matched only by Zone-II model predictions with pocket mass below $0.5\times10^{-3}$ $M_\odot$ in the ST case. Interestingly, by extending the pocket mass to its maximum in Torino postprocess AGB models, we found that Zone-II_p16 model in the D7.5 case (Figure 10h) also predicts $\delta(^{138}\text{Ba}/^{136}\text{Ba})$ to lie below −400‰ (−420‰ as its minimum) for the first several TPs during the carbon-rich phase, while its predicted $\delta(^{88}\text{Sr}/^{86}\text{Sr})$ range is more compatible with the values observed in the three grains with $\delta(^{138}\text{Ba}/^{136}\text{Ba})$ below −400‰. Thus, these extremely negative $\delta(^{138}\text{Ba}/^{136}\text{Ba})$ values observed in the mainstream SiC grains might be an indicator of parent AGB stars with extremely large $^{13}$C-pockets. Nevertheless, mainstream SiC grains with $\delta(^{138}\text{Ba}/^{136}\text{Ba}) < -400‰$ are rare (< 10%), as shown by both Liu et al. (2014a) and this study.

*4.4. Mainstream SiC Grains Without s-Process Isotopic Signature*

Six of the 61 mainstream SiC grains measured in this study show close-to-solar strontium and barium isotopic compositions (open squares in Figure 3). As discussed in Section 3.2, solar system contamination on grain surfaces is unlikely in these six grains. Figure 13 shows that $^{13}$C mass fractions lower than the D12 case in the Three-zone model are required to reach close-to-solar $\delta(^{84}\text{Sr}/^{86}\text{Sr})$ values, which corresponds to a total $^{13}$C mass of less than $1.6\times10^{-7}$ $M_\odot$ within the $^{13}$C-pocket. As shown in Table 6, the majority of mainstream SiC grains can be explained by a total $^{13}$C mass of $\sim(6-30)\times10^{-7}$ $M_\odot$. Thus, it is highly likely that these six mainstream grains came from parent AGB stars with inefficient *s*-process nucleosynthesis during the carbon-rich period of their AGB phase. Inefficient *s*-process nucleosynthesis, *i.e.*, a low mean neutron exposure, could be caused by high initial stellar metallicity (high iron seed abundance), or by a low $^{13}$C abundance and/or high $^{14}$N abundance in the $^{13}$C-pocket, perhaps due to *e.g.*, extremely high rotation speeds of parent AGB stars. In fact, the $\delta(^{29,30}\text{Si}/^{28}\text{Si})$ values of these six mainstream SiC grains vary from −100‰ to 150‰ (Table 1), spanning the whole range of values observed in mainstream SiC grains (Table 1 and Figure 10a). Consistent with our conclusion in Section 4.1.2, this observation in mainstream SiC grains without *s*-process isotopic signatures excludes initial stellar metallicity as the dominant factor in reducing the mean neutron exposure in AGB stars.

### 4.5. Implications for $^{13}$C-Pocket Formation

#### 4.5.1. Independent constraints on the $^{13}$C-pocket mass and the $^{13}$C mass fraction

Strontium and barium isotopic compositions in acid-cleaned mainstream SiC grains from this study point towards the common existence of parent AGB stars with larger $^{13}$C-pockets (1−2×10$^{−3}$ $M_\odot$) and lower $^{13}$C mass fractions than previously assumed. In previous attempts of *s*-process nucleosynthesis models in AGB stars to reproduce observations and grain data (*e.g.*, Gallino et al. 1998, Busso et al. 2001, Barzyk et al. 2007), the effect of the $^{13}$C-pocket mass was never distinguished from that of the $^{13}$C mass fraction due to the lack of a means of separating the two effects. Instead, investigation of $^{13}$C-pocket formation was mainly based on matching the total mass of $^{13}$C nuclei within the $^{13}$C-pocket, which actually depends on combined effects of three variables. For instance, AGB model calculations with larger $^{13}$C-pockets overpredict astronomical observations of [hs/ls] in AGB stars, where [hs/ls] is the log of the ratio of heavy-*s* (barium-peak) to light-*s* (zirconium-peak) elements divided by the same ratio in the solar system (see, *e.g.*, Table 1 of Herwig et al. 2003). In fact, the [hs/ls] predictions could be easily lowered by reducing the $^{13}$C mass fraction within the larger $^{13}$C-pocket (physically corresponding to *e.g.*, a higher rotation speed) because the final *s*-process production does not only depend on the total $^{13}$C mass, but is the result of $^{13}$C pocket mass, $^{13}$C mass fraction and, to a lesser extent, the $^{13}$C profile. By investigating the correlated $\delta(^{88}Sr/^{86}Sr)$ and $\delta(^{138}Ba/^{136}Ba)$ values in mainstream SiC grains, we have resolved the effects and derived independent constraints on the two parameters of the $^{13}$C-pockets in parent AGB stars. As shown in Figure 9, even if large $^{13}$C-pockets (up to 8×10$^{−3}$ $M_\odot$ pocket mass) form in AGB stars, lower-than-D4.5 $^{13}$C mass fractions are required for 0.5 $Z_\odot$ model predictions (lower-than-D2.2 $^{13}$C mass fractions for $Z_\odot$ model predictions) to match the mainstream SiC grain data. Recently, Maiorca et al. (2012) have also proposed large $^{13}$C-pockets (similar to the pocket mass of Three-zone_p4) in AGB stars with stellar mass lower than 1.5 $M_\odot$ to explain the abundances of neutron-rich elements in young open clusters. Quantitative studies of the effect of magnetic fields and rotation on $^{13}$C-pocket formation may provide additional information on the $^{13}$C-pocket structure, which can be confirmed by comparing the AGB model predictions with the grain data from this study.

#### 4.5.2. Combined effect of convective overshoot and rotation

A variety of physical mechanisms have been proposed for the $^{13}$C-pocket formation in AGB stars, including convective overshoot, gravity waves, rotation, and magnetic fields.

Overshoot has been proposed repeatedly (Straniero et al. 1995; Freytag et al. 1996; Herwig et al. 1997; Cristallo et al. 2001). In the framework of the mixing length theory, fluid motions come to a full stop at the boundary between convective and radiative regions. This transition is normally mild. However, during TDU episodes, the drop of the convective velocities is abrupt due to the sudden change in the opacity encountered from the hydrogen-rich (and opaque) convective envelope and the underlying radiative helium-rich (and less opaque) helium intershell. Thus, during this specific phase, the lower boundary of the convective envelope is unstable and any perturbation may cause further mixing (see Straniero et al. 2014). Instead of stopping, it is more likely that an eddy crosses the boundary and overshoots into the stable helium intershell. In this way, hydrogen is brought into the helium intershell with a profile that decreases with depth. It has been found that overshoot alone cannot form $^{13}$C-pockets with pocket mass higher than $1\times10^{-3}$ $M_\odot$ (*e.g.*, Herwig et al. 1997; Cristallo et al. 2009) and that the extension in mass of the pocket decreases pulse-by-pulse along the AGB track (Cristallo et al. 2011). This upper limit of the pocket mass was historically adopted for the Three-zone $^{13}$C-pocket in Torino postprocess AGB model calculations.

As convective overshoot is not expected to be a stochastic process (Herwig et al. 2003), variation in the $^{13}$C mass fraction among parent AGB stars of mainstream SiC grains could more likely be explained by different angular velocity gradients at the core-envelope interface in rotating AGB stars with different initial rotation speeds possibly interacting with magnetic fields generated by the motions. Piersanti et al. (2013) have recently investigated the combined effect of the convective overshoot and the rotational motion on the $^{13}$C-pocket evolution in 2 $M_\odot$, $Z_\odot$ AGB stars with rotation speeds ranging from 10 to 120 km s$^{-1}$. While the formation of the $^{13}$C-pocket is entirely ascribed to the opacity-induced overshoot, rotation has important effects on its evolution during the interpulse phase (see Figure 3 of Piersanti et al. 2013). In order to follow the nucleosynthesis of heavy element nuclides in rotating models in a more realistic way than the postprocess AGB models, FRUITY models are computed with a full nuclear network directly coupled to the physical evolution of an AGB star accounting for the complex interplay between the nuclear burning and the rather inefficient mixing induced by rotation.

FRUITY models adopt the solar abundance distribution from Lodders (2003), which has a slightly lower initial solar metallicity ($Z = 0.0138$, the mass fraction of elements heavier than helium in the solar system 4.56 Ga ago) compared to $Z = 0.0153$ in Lodders et al. (2009) adopted in the Torino models. As the difference in the initial iron content (the seed of the *s*-process) is less than 5%, it has negligible consequences on *s*-process nucleosynthesis calculations in the two models. The $^{22}$Ne($\alpha,n$)$^{25}$Mg rate at AGB stellar temperature is from Jaeger et al. (2001), which is similar to the ½×K94 rate adopted in Torino postprocess AGB models.

In the sections below, we compare recent rotating FRUITY models with our grain data. As explained earlier, while postprocess models provide a fundamental method to test and address the characteristics of the internal structure of the $^{13}$C-pocket, full evolutionary models are crucial to provide a more self-consistent approach (by including possible physical explanations) for the most promising tests. Note that magnetic effects are not included in these models.

Comparison of our grain data with FRUITY rotating model calculations is shown in Figure 14 for 2 $M_\odot$ AGB stars with close-to-solar metallicities. Model predictions for $\delta(^{88}$Sr/$^{86}$Sr) strongly decrease with increasing rotation speed; the final $\delta(^{138}$Ba/$^{136}$Ba) predictions barely vary with varying rotation speeds. The spread of the $\delta(^{135}$Ba/$^{136}$Ba) and $\delta(^{138}$Ba/$^{136}$Ba) predictions during the carbon-rich phase increases with increasing rotation speed (Figures 14a,b). As shown by Piersanti et al. (2013), the GSF instability is active in the upper region of the $^{13}$C-pocket in rotating AGB stars and contaminates the $^{13}$C-pocket with the major neutron poison $^{14}$N (Figure 2 of Piersanti et al. 2013). As a result, mixing caused by the GSF instability reduces the effective $^{13}$C mass fraction within the $^{13}$C-pocket, and spreads the $^{13}$C nuclei over a larger zone. In general, the larger the initial rotation speed, the higher the GSF efficiency, and in turn, the larger the unstable zone. Thus, the lower neutron-to-seed ratios within the $^{13}$C-pocket caused by $^{14}$N contamination decreases the $\delta(^{88}$Sr/$^{86}$Sr) predictions, which better agrees with the mainstream grain data in Figures 14a,b. This dilution scenario is supported by Figures 9 and 11: as the pocket mass increases, lower $^{13}$C mass fractions are required to match the isotopic compositions of mainstream SiC grains.

We also investigated the combined effect of metallicity and rotation by computing a new set of 2 $M_\odot$ rotating models with metallicities at $Z = 0.72\ Z_\odot$ and $Z = 1.45\ Z_\odot$ ($Z_\odot = 0.0138$), in comparison to the grain data from this study in Figures 14c−f. As discussed in Section 4.1.1, an increase in the metallicity decreases AGB model predictions for $\delta(^{88}$Sr/$^{86}$Sr), and the predictions therefore agree better with the grain data. However, the supersolar metallicity models cannot match the *s*-process overproduction factors observed in the grains, *i.e.*, the range of $\delta(^{135}$Ba/$^{136}$Ba) values. The situation is reversed for the subsolar metallicity cases, where the range of $\delta(^{135}Ba/^{136}Ba)$ values can be matched, but the $\delta(^{88}Sr/^{86}Sr)$ predictions are too high. In summary, it seems that change in metallicity (in either direction) does not lead to a significantly better fit of FRUITY models to the grain data. In fact, no FRUITY model can cover the grain data region with both low $\delta(^{88}Sr/^{86}Sr)$ and $\delta(^{135}Ba/^{136}Ba)$ values.

### 4.5.3. Constraints on the $^{13}C$-pocket formation

According to Piersanti et al. (2013), although rotation could significantly improve the match between the FRUITY predictions and the grain data, apparently convective overshoot and rotation alone cannot provide the right $^{13}C$-pocket to account for the *s*-process isotopic signatures observed in mainstream SiC grains. Cristallo et al. (2015) have recently found larger $^{13}C$-pockets than those characterizing nonrotating FRUITY models, where a different boundary condition for the penetration of convective elements is assumed during TDU episodes. This way, a deeper penetration of protons has been obtained with a very low mixing efficiency. As a result, the new $^{13}C$-pockets present an extended tail with lower $^{13}C$ mass fractions (hereafter "tailed cases"). The model predictions are shown in Figure 15 for a 2 $M_\odot$, 0.72 $Z_\odot$ star. While barium isotopes are practically unchanged, a decrease of $\delta(^{88}Sr/^{86}Sr)$ is found in the "tailed" models. When this new treatment of the $^{13}C$-pocket is coupled with rotation (initial rotation speed = 30 km s$^{-1}$), a further decrease (−100‰ lower) in the $\delta(^{88}Sr/^{86}Sr)$ predictions is obtained , but the predictions still lie outside of the grain data regime.

We also evaluated the effect of the uncertainties in the strontium isotope MACSs on $\delta(^{88}Sr/^{86}Sr)$ predictions. We mentioned earlier that the FRUITY models are based on the MACSs recommended by Bao et al. (2000). The values between the provided energies are interpolated based on the $1/v_T$ rule. The $^{88}Sr$ MACS, however, peaks at 8 keV, thus deviating from such a rule in the 5–10 keV energy range (Figure 1). We investigated the strontium MACS uncertainty by adopting the lower limit $^{86}Sr$ MACS and the upper limit $^{88}Sr$ MACS in the KADoNiS v1.0 database in the 0 km s$^{-1}$ and 30 km s$^{-1}$ FRUITY rotating AGB models. Compared to the corresponding models, significant decreases in $\delta(^{88}Sr/^{86}Sr)$ (~300‰ in the 0 km s$^{-1}$ case, and ~200‰ in the 30 km s$^{-1}$ case) are found in the model predictions with the new strontium MACSs, which lie closer to the grain data regime.

Another possibility to reach a better match with the grain data in Figure 14 is the consideration of mixing caused by other process(es), such as gravity waves excited by turbulent motions near the base of the convective envelope (Denissenkov & Tout 2003) or magnetic fields (Busso et al. 2012). In fact, it cannot be excluded that various mixing processes may operate simultaneously, interacting with each other. For instance, gravity waves are expected to work during TDU episodes and magnetic fields develop in rotating stars. One or more other mixing processes in addition to overshoot and rotation may be invoked to shape the large $^{13}$C-pockets with low $^{13}$C abundances suggested by the grain data from this study.

## 5. CONCLUSIONS

We measured correlated strontium and barium isotope ratios in 61 acid-cleaned mainstream SiC grains from Murchison. Although varying initial stellar metallicities of parent AGB stars could help to explain the spread of mainstream grain data for strontium and barium isotopes, we did not observe any clear correlation between the implied initial stellar metallicity ($^{29,30}$Si/$^{28}$Si ratios) and the abundances of neutron-magic nuclei ($^{88}$Sr and $^{138}$Ba) in mainstream SiC grains.

Instead, the spread of the strontium and barium isotopic data from this study is more likely to be a natural consequence of varying $^{13}$C-pockets shaped by different degrees of mixing in parent AGB stars. The order of importance in determining the range of *s*-process isotopic compositions observed in mainstream SiC grains is: $^{13}$C mass fraction, $^{13}$C-pocket mass, and $^{13}$C profile. We found that AGB model predictions for $\delta(^{88}\text{Sr}/^{86}\text{Sr})$ are positively correlated with the $^{13}$C mass fraction within the $^{13}$C-pocket. In addition, the constrained mass fractions from previous studies (Barzyk et al. 2007) overpredict $\delta(^{88}\text{Sr}/^{86}\text{Sr})$ values for mainstream SiC grains. By investigating lower $^{13}$C mass fractions, we found that previously assumed $^{13}$C-pockets with masses below $1\times10^{-3}$ $M_\odot$ are insufficient to match the strontium and barium isotopic signatures observed in mainstream SiC grains. A larger $^{13}$C-pocket, $\geq 1\times10^{-3}$ $M_\odot$, with a range of relatively low $^{13}$C mass fractions is required to explain the mainstream grain data for these isotopes. In addition, variation in the neutron-to-seed ratio of about a factor of two to three within a fixed-mass $^{13}$C-pocket is required to explain the range of observed isotopic compositions for both decreasing-with-depth and flat $^{13}$C profiles.

Finally, we showed that the inclusion of rotation and a different treatment of the lower boundary of the $^{13}$C-pocket in FRUITY models can improve the fit to the grain data, even if these models still cannot explain some of the grain data. We also evaluated the effects of nuclear uncertainties. We point out the possibility that various mixing processes may work together in shaping the $^{13}$C-pockets in AGB stars, resulting in both large $^{13}$C-pockets and a range of relatively low $^{13}$C mass fractions within the $^{13}$C-pockets.

Acknowledgements: We thank Prof. Ernst Zinner for his help with the NanoSIMS measurement. We also thank Drs. Oscar Straniero and Luciano Piersanti for many enlightening discussions on stellar modeling. The CHARISMA instrument at Argonne National Laboratory is supported in part by the U.S. Department of Energy, Office of Science, Materials Sciences and Engineering Division. NL acknowledges the NASA Earth and Space Sciences Fellowship program (NNX11AN63H) for support. NL and AMD acknowledge the NASA Cosmochemistry program (NNX09AG39G) for support. MRS acknowledges support from the NASA Cosmochemistry program (NNH08AI81I). SB acknowledges financial support from the Joint Institute for Nuclear Astrophysics (JINA, University of Notre Dame, USA) and from Karlsruhe Institute of Technology (KIT, Karlsruhe, Germany). SC acknowledges the Italian Grants RBFR08549F-002 (FIRB 2008 program) and PRIN-MIUR 2012 "Nucleosynthesis in AGB stars: an integrated approach" project (20128PCN59) for support. Part of the Torino model numerical calculations has been sustained by B$^2$FH Association (http://www.b2fh.org/).

FIGURE CAPTIONS

Figure 1  Comparison of the MACS values of $^{86}$Sr, $^{87}$Sr, $^{88}$Sr, and $^{85}$Kr in KADoNiS v1.0 with those in KADoNiS v0.3.

Figure 2  The krypton to strontium section of the chart of the nuclides. Solar system abundances are shown as atom percentages for stable isotopes (solid squares); laboratory half-lives at room temperature are shown for unstable isotopes (dotted squares). Pure *s*-process nuclides are outlined with thick black squares. The main Path (1) of the *s*-process is shown with thick arrows and the main Path (2) with thick dash-dotted arrows; alternative paths (due to branch points) are indicated with thin arrows. Neutron-magic nuclides lie on the vertical yellow band at $N = 82$. The decay of long-lived nuclide $^{87}$Rb ($t_{1/2} = 49.2$ Ga) to $^{87}$Sr is indicated by a dashed arrow.

Figure 3  Three-isotope plots of δ($^{137}$Ba/$^{136}$Ba) versus δ($^{135}$Ba/$^{136}$Ba), and δ($^{87}$Sr/$^{86}$Sr) versus δ($^{84}$Sr/$^{86}$Sr) for the 61 grains in this study. Six mainstream grains with close-to-solar isotopic compositions are shown as open squares (highlighted in Table 1). Note that barium isotope ratios are not available for one of the six grains (G189). Unclassified grains (triangles) are well within the range of mainstream grains and are therefore grouped as mainstream and shown as black dots hereafter. Uncertainties are ±2σ. Dotted lines represent solar isotope ratios.

Figure 4  Three-isotope plots of δ($^{137}$Ba/$^{136}$Ba) versus δ($^{135}$Ba/$^{136}$Ba), and δ($^{87}$Sr/$^{86}$Sr) versus δ($^{84}$Sr/$^{86}$Sr) for comparison with single mainstream grain data in the literature.

Figure 5  Three-isotope plot of δ($^{87}$Sr/$^{86}$Sr) versus δ($^{84}$Sr/$^{86}$Sr) for comparison of 55 single mainstream SiC grain data with available δ($^{84}$Sr/$^{86}$Sr) values from this study with AGB model predictions for a 2 $M_\odot$, 0.5 $Z_\odot$ AGB star.

Figure 6  Three-isotope plots of δ($^{137}$Ba/$^{136}$Ba) versus δ($^{135}$Ba/$^{136}$Ba), and δ($^{87}$Sr/$^{86}$Sr) versus δ($^{84}$Sr/$^{86}$Sr). Linear regression lines (black solid lines) of single grain data with 95% confidence (shown as grey region) are compared with those of SiC aggregates (red dashed lines). Data sources: barium isotope data in single SiC grains from this study and Liu et al. (2014a); barium isotope data in SiC aggregates from Ott & Begemann (1990a), Prombo et al. (1993), and Zinner et al. (1991); strontium isotope data in single SiC grains from this study; strontium isotope data in SiC aggregates from Ott & Begemann (1990b) and Podosek et al. (2004).

Figure 7  Three-isotope plots of δ($^{88}$Sr/$^{86}$Sr) versus δ($^{84}$Sr/$^{86}$Sr). Forty-nine single mainstream SiC grain data are compared to Three-zone models for AGB stars with a range of mass and metallicity. Note that six grains with close-to-solar isotope ratios are not shown. The entire evolution of the AGB envelope composition is shown, but symbols are plotted only when C > O.

Figure 8  Plots of: (a) δ($^{29}$Si/$^{28}$Si) versus δ($^{29}$Si/$^{28}$Si), (b) δ($^{88}$Sr/$^{86}$Sr) versus δ($^{29}$Si/$^{28}$Si), (c) δ($^{138}$Ba/$^{136}$Ba) versus δ($^{29}$Si/$^{28}$Si), (d) δ($^{88}$Sr/$^{86}$Sr) versus δ($^{30}$Si/$^{28}$Si), and (e) δ($^{138}$Ba/$^{136}$Ba) versus δ($^{30}$Si/$^{28}$Si). The mainstream SiC grain data are from Liu et al. (2014a) and this study (black dots) and presolar grain database (Hynes & Gyngard 2009). Best-fit lines with 95% confidence bands are shown if there seems to be a linear correlation.

Figure 9   Four-isotope plots of δ($^{88}$Sr/$^{86}$Sr) versus δ($^{138}$Ba/$^{136}$Ba). Three-zone and Zone-II model predictions with varying $^{13}$C-pocket masses (~0.5−8 × 10$^{−3}$ $M_\odot$) for a 2 $M_\odot$, 0.5 $Z_\odot$ AGB star are compared to the 47 mainstream SiC grain data from this study with available correlated δ($^{88}$Sr/$^{86}$Sr) and δ($^{138}$Ba/$^{136}$Ba) values. The K94 $^{22}$Ne($\alpha$,$n$)$^{25}$Mg rate is adopted in all model calculations.

Figure 10   Four-isotope plots of δ($^{88}$Sr/$^{86}$Sr) versus δ($^{138}$Ba/$^{136}$Ba). The same set of grains as in Figure 9 is compared to Three-zone model predictions with varying nuclear inputs for a 2 $M_\odot$, 0.5 $Z_\odot$ AGB star. KAD refers to the KADoNiS v1.0 database.

Figure 11   Four-isotope plots of δ($^{88}$Sr/$^{86}$Sr) versus δ($^{135}$Ba/$^{136}$Ba). The models constrained in Table 4 are compared to the 47 mainstream SiC grain data from this study with available correlated δ($^{88}$Sr/$^{86}$Sr) and δ($^{135}$Ba/$^{136}$Ba) values. The K94 $^{22}$Ne($\alpha$,$n$)$^{25}$Mg rate is adopted in all model calculations.

Figure 12   Plots of δ($^{88}$Sr/$^{86}$Sr) versus δ($^{84}$Sr/$^{86}$Sr). The selected mainstream SiC grains (see text for details) are compared to the constrained models in Table 6. The K94 rate is adopted in all the model calculations.

Figure 13   Three-isotope plots of δ($^{88}$Sr/$^{86}$Sr) versus δ($^{84}$Sr/$^{86}$Sr), and δ($^{138}$Ba/$^{136}$Ba) versus δ($^{135}$Ba/$^{136}$Ba). The Three-zone model predictions in the D7.5 and D12 cases for a 2 $M_\odot$, 0.5 $Z_\odot$ AGB star are compared to the six mainstream SiC grains (highlighted in Table 1) with close-to-solar strontium and/or barium isotopic compositions.

Figure 14   Four-isotope plots of δ($^{88}$Sr/$^{86}$Sr) versus δ($^{138}$Ba/$^{136}$Ba) in (a,c,e) and δ($^{88}$Sr/$^{86}$Sr) versus δ($^{135}$Ba/$^{136}$Ba) in (b,d,f). The mainstream SiC grain data from this study are compared to FRUITY rotating model predictions for 2 $M_\odot$ AGB stars with metallicities at 0.72 $Z_\odot$, $Z_\odot$, and 1.45 $Z_\odot$ by Piersanti et al. (2013).

Figure 15   Four-isotope plots of δ($^{88}$Sr/$^{86}$Sr) versus δ($^{138}$Ba/$^{136}$Ba) and δ($^{88}$Sr/$^{86}$Sr) versus δ($^{135}$Ba/$^{136}$Ba). Test results of (1) a "tailed" $^{13}$C-pocket by Cristallo et al. (2015) and (2) the KADoNiS v1.0 (KAD) strontium MACSs in 2 $M_\odot$, 0.72 $Z_\odot$ FRUITY rotating models are compared with the grain data from this study.

Table 1. Carbon, Silicon, Strontium, and Barium Grain Data

| Grains | Type | $^{12}C/^{13}C$ | $\delta^{29}Si$ (‰) | $\delta^{30}Si$ (‰) | $\delta^{84}Sr$ (‰) | $\delta^{87}Sr$ (‰) | $\delta^{88}Sr$ (‰) | $\delta^{130+132}Ba$ (‰) | $\delta^{134}Ba$ (‰) | $\delta^{135}Ba$ (‰) | $\delta^{137}Ba$ (‰) | $\delta^{138}Ba$ (‰) |
|---|---|---|---|---|---|---|---|---|---|---|---|---|
| G3 | M | 81±0.8 | −1±9 | −10±11 | −721±170 | −165±186 | −170±129 | | 228±579 | −556±218 | −352±276 | −93±267 |
| G5 | M | 69±0.9 | 18±18 | −43±21 | <−879 | −214±167 | −143±124 | −792±158 | −48±164 | −765±49 | −520±73 | −359±64 |
| G6 | M | 55±0.4 | 6±10 | −5±12 | −905±63 | 152±193 | 94±138 | −872±60 | 308±312 | −742±80 | −527±113 | −391±95 |
| G17 | M | 52±0.2 | 22±9 | 5±10 | −868±68 | −100±96 | −150±65 | −835±106 | 382±261 | −666±76 | −408±107 | −237±95 |
| G21 | M | 64±0.7 | 22±7 | 12±8 | −692±116 | 15±95 | −82±63 | | 224±349 | −524±138 | −353±167 | −236±138 |
| G26 | M | 49±0.5 | 142±8 | 91±9 | −941±33 | −113±127 | −177±84 | | 113±119 | −539±49 | −389±58 | −269±48 |
| G27 | M | 53±0.2 | −30±5 | 23±5 | | | | −642±244 | 192±348 | −548±135 | −454±149 | −271±133 |
| G29 | M | 62±0.4 | 73±7 | 58±7 | −771±143 | −78±115 | −131±77 | | 53±341 | −648±122 | −494±149 | −323±132 |
| G30 | M | 28±0.1 | 69±6 | 64±5 | −840±81 | 72±107 | −56±70 | | −142±144 | −548±69 | −368±84 | −276±67 |
| **G32** | **M** | **73±1.1** | **−112±11** | **−68±14** | **179±605** | **205±299** | **173±221** | | **189±416** | **−189±242** | **−36±276** | **39±223** |
| G33 | M | 60±0.4 | −111±6 | −58±7 | −742±115 | −22±112 | −118±73 | | 3±163 | −510±74 | −351±88 | −203±75 |
| G41 | M | 91±0.8 | 93±12 | 33±13 | <−747 | 196±438 | −150±230 | | −39±248 | −752±73 | −542±87 | −316±91 |
| G42 | M | 36±0.2 | 188±12 | 99±13 | −828±174 | 1±184 | −189±105 | | 81±326 | −364±163 | −282±149 | −220±126 |
| G54 | M | 50±0.5 | 120±9 | 75±10 | <−535 | 59±244 | −30±158 | | | | | |
| G56 | M | 20±0.2 | 78±7 | 55±8 | −783±311 | 28±364 | −123±219 | | | | | |
| **G64** | **M** | **52±0.7** | **−28±14** | **−2±17** | **−297±268** | **−168±164** | **−218±103** | | **335±251** | **−156±133** | **−132±115** | **−75±99** |
| G99 | M | 46±0.3 | −30±6 | 1±17 | −987±13 | −149±263 | 227±289 | | 125±84 | −834±35 | −511±56 | −519±55 |
| G100 | M | 86±1.7 | 252±114 | 98±129 | −910±28 | 29±175 | −229±108 | | -70±174 | −710±114 | −575±120 | −357±167 |
| G104 | M | 70±2.1 | 46±36 | 12±42 | −955±16 | 65±155 | −41±115 | | 190±182 | −369±169 | −258±162 | −104±205 |
| G110 | M | 89±0.6 | 17±6 | 15±6 | <−795 | 47±136 | −92±84 | | 188±152 | −717±41 | −461±52 | −330±48 |
| G168 | M | 62±3.3 | 48±77 | −26±90 | −966±41 | 31±496 | −9±388 | | −347±199 | −486±232 | −350±230 | −349±239 |
| G185 | M | 90±0.9 | 6±6 | 0±8 | | 119±354 | −311±160 | | −74±386 | −599±157 | −305±189 | −216±165 |
| **G189** | **M** | **47±0.3** | **155±7** | **136±6** | **191±381** | **21±138** | **−27±95** | | | | | |
| G193 | M | 53±0.3 | 193±7 | 138±7 | −849±92 | −49±99 | −144±64 | | −104±132 | −527±128 | −246±145 | −290±115 |
| G196 | M | 82±0.9 | 104±8 | 80±8 | | 129±220 | 77±155 | | 76±107 | −456±101 | −263±103 | −231±89 |
| **G214** | **M** | **29±0.2** | **80±8** | **105±8** | **−164±166** | **4±129** | **−82±99** | | **84±426** | **−48±304** | **9±279** | **−34±204** |
| **G219** | **M** | **77±0.4** | **−9±5** | **36±5** | **169±231** | **−9±84** | **−59±50** | | **150±202** | **30±131** | **32±116** | **−22±90** |
| G229 | M | 92±1.3 | 125±9 | 100±11 | <−775 | −93±276 | −269±125 | | 93±348 | −371±169 | −334±149 | −298±177 |
| G235 | M | 50±0.5 | 35±7 | 54±8 | <−738 | −88±151 | −35±88 | | 185±167 | −724±45 | −459±58 | −404±69 |
| G266 | M | 63±0.7 | 3±7 | 17±8 | −855±51 | −51±87 | −145±38 | | 429±420 | −456±177 | −334±153 | −324±59 |
| G299 | M | 83±0.4 | 38±6 | 27±5 | −958±21 | −172±162 | −153±105 | | 32±196 | −703±122 | −379±168 | −250±150 |
| G312 | M | 49±0.3 | 134±6 | 126±6 | −961±33 | −38±234 | −155±136 | | 84±165 | −595±120 | −281±152 | −317±112 |
| G328 | M | 46±0.2 | 34±6 | 43±6 | <−722 | 23±161 | −129±117 | −796±191 | −227±277 | −550±145 | −326±163 | −279±124 |
| G334 | M | 50±0.2 | 85±6 | 97±5 | −909±68 | 659±514 | −13±239 | | −26±238 | −556±135 | −499±130 | −510±88 |
| G335 | M | 48±0.3 | −13±7 | 21±7 | −993±4 | −1±48 | 45±35 | −909±49 | 287±81 | −734±17 | −469±24 | −363±22 |
| G367 | M | 92±0.9 | −28±6 | 11±8 | <−871 | 42±231 | 19±130 | | 53±214 | −729±62 | −428±83 | −85±142 |
| G417 | M | 67±0.7 | 79±11 | 60±14 | −910±52 | 15±73 | −79±48 | | 62±118 | −667±82 | −365±102 | −349±85 |
| G418 | M | 55±0.4 | 75±8 | 49±10 | −912±43 | −24±78 | −179±47 | | 81±102 | −580±82 | −339±91 | −313±77 |
| G421 | M | 75±0.9 | 4±23 | 5±28 | −924±58 | −24±90 | −99±60 | | 87±130 | −663±90 | −389±108 | −474±76 |
| **G422** | **M** | **51±0.4** | **51±8** | **55±10** | **−282±290** | **−30±135** | **−143±86** | | **165±142** | **−194±168** | **−131±148** | **−152±123** |
| G441 | M | 67±0.4 | 81±11 | 47±12 | −992±4 | −11±56 | −104±34 | | 20±91 | −713±55 | −446±72 | −354±61 |
| G445 | M | 45±0.3 | 46±13 | 25±15 | <−771 | −30±60 | −172±43 | | -92±187 | −538±90 | −336±99 | −308±73 |
| G446 | M | 82±0.5 | 23±9 | 9±10 | | 10±429 | −127±314 | | 333±447 | −589±157 | −353±181 | −65±182 |
| G449 | M | 57±0.6 | 117±13 | 113±16 | | −78±121 | −83±99 | | 262±250 | −613±87 | −386±100 | −266±83 |
| G450 | M | 47±0.4 | 208±25 | 141±30 | <−899 | −7±302 | −262±151 | | 104±199 | −459±173 | −326±173 | −400±119 |
| G453 | M | 52±0.3 | 123±8 | 125±10 | −986±10 | 6±135 | −44±85 | −813±140 | 344±126 | −705±65 | −401±88 | −339±72 |
| G454 | M | 51±0.4 | 140±8 | 142±10 | −991±3 | 31±57 | −113±33 | | 118±84 | −563±62 | −291±75 | −278±59 |
| G475 | M | 49±0.3 | 158±9 | 143±12 | −984±12 | 44±206 | 140±150 | | 135±87 | −645±56 | −413±67 | −357±54 |
| G477 | M | 90±0.6 | 64±8 | 73±10 | −959±22 | 100±221 | 5±138 | −688±219 | 17±96 | −764±52 | −389±82 | −296±70 |
| G478 | M | 71±0.5 | 38±10 | 75±13 | −972±15 | −218±191 | −111±135 | | 21±215 | −686±139 | −356±190 | −266±162 |
| G71 | U[a] | | | | <−804 | 0±250 | −11±180 | | 36±159 | −724±49 | −488±71 | −357±59 |
| G103 | U | | | | <−896 | 147±203 | 61±157 | | | | | |
| G191 | U | | | | −852±35 | 30±42 | −119±26 | | 24±114 | −548±99 | −350±103 | −303±90 |
| G211 | U | | | | −902±39 | −5±193 | 20±135 | | | | | |
| G222 | U | | | | −864±128 | 496±562 | −87±218 | | −8±272 | −800±68 | −534±95 | −176±171 |
| G226 | U | | | | | −53±234 | −35±134 | | 359±334 | −632±99 | −373±118 | −288±150 |
| G304 | U | | | | −990±9 | −33±120 | −152±70 | | 261±91 | −699±49 | −429±63 | −341±54 |
| G427 | U | | | | <−898 | −190±158 | 105±162 | | | | | |
| G439 | U | | | | −991±6 | −6±84 | −196±45 | | | | | |
| G444 | U | | | | | −7±199 | −146±144 | | −203±332 | −568±167 | −331±191 | −397±124 |
| G492 | U | | | | <−620 | 4±174 | −94±113 | | | | | |

**Notes**: Uncertainties are given as 2σ. Mainstream grains with close-to-solar isotopic compositions are highlighted in bold.

[a]: U stands for unclassified.

Table 2. The Parameters of the $^{13}$C-Pockets Adopted in Torino Postprocess AGB Models in the ST Case.

| Models | Zone I | | | Zone II | | | Zone III | | |
|---|---|---|---|---|---|---|---|---|---|
| | $M$ ($10^{-4} M_\odot$) | $X(^{13}C)$ ($10^{-3}$) | $X(^{14}N)$ ($10^{-4}$) | $M$ ($10^{-4} M_\odot$) | $X(^{13}C)$ ($10^{-3}$) | $X(^{14}N)$ ($10^{-4}$) | $M$ ($10^{-4} M_\odot$) | $X(^{13}C)$ ($10^{-3}$) | $X(^{14}N)$ ($10^{-4}$) |
| Three-zone_d2.5 | 1.600 | 3.20 | 1.07 | 2.120 | 6.80 | 2.08 | 0.030 | 16.00 | 2.08 |
| Three-zone | 4.000 | 3.20 | 1.07 | 5.300 | 6.80 | 2.08 | 0.075 | 16.00 | 2.08 |
| Three-zone_P2 | 8.000 | 3.20 | 1.07 | 10.600 | 6.80 | 2.08 | 0.150 | 16.00 | 2.08 |
| Three-zone_P8 | 32.000 | 3.20 | 1.07 | 40.600 | 6.80 | 2.08 | 0.600 | 16.00 | 2.08 |
| Zone-II | | | | 2.120 | 6.80 | 2.08 | | | |
| Zone-II_p2 | | | | 5.300 | 6.80 | 2.08 | | | |
| Zone-II_p4 | | | | 10.600 | 6.80 | 2.08 | | | |
| Zone-II_p16 | | | | 40.600 | 6.80 | 2.08 | | | |

Note: For Torino postprocess model calculations in other cases (e.g., D1.5), the pocket mass ($M$) remains the same while the $^{13}$C mass fraction is that in the ST case divided by the case number.

Table 3. Strontium Isotope MACS values (mb) recommended by Koehler et al. (2000) (Koe 2000), KADoNiS v0.3 (KAD v0.3), and KADoNiS v1.0 (KAD v1.0).

| | Energy (keV) | 5 | 8 | 10 | 15 | 20 | 25 | 30 | 1σ err (%) @ 30 keV |
|---|---|---|---|---|---|---|---|---|---|
| $^{86}$Sr | KAD v0.3 | 211 | | 134 | 100 | 86 | 71 | 64±3 | 4.69 |
| | KAD v1.0 | 219±24 | | 124±17 | 94.1±11.5 | 79.1±8.1 | 70±6.1 | 63.4±4.7 | 7.41 |
| $^{87}$Sr | KAD v0.3 | 345 | | 207 | 155 | 127 | 110 | 92±4 | 4.35 |
| | KAD v1.0 | 286±13 | | 195±10 | 151±8 | 125±7 | 107±7 | 94.6±6.2 | 6.55 |
| $^{88}$Sr | Koe 2000 | 9.19±0.30 | 10.70±0.34 | 10.33±0.33 | 8.88±0.27 | 7.60±0.23 | 6.68±0.20 | 6.01±0.17 | 2.83 |
| | KAD v0.3 | 10.88 | | 11.86 | 9.88 | 8.21 | 7.02 | 6.13±0.11 | 1.79 |
| | KAD v1.0 | 9.42±0.31 | 10.97±0.35 | 10.59±0.35 | 9.10±0.28 | 7.79±0.24 | 6.85±0.20 | 6.16±0.17 | 2.76 |

Table 4. Constraints on $(1-2) \times 10^{-3}\, M_\odot$ $^{13}$C-Pocket Using Correlated $\delta(^{88}Sr/^{86}Sr)$ and $\delta(^{138}Ba/^{136}Ba)$ Grain Data.

| Models | K94 KAD $^{86}$Sr MACS | |
|---|---|---|
| | $^{13}$C-Pocket Mass $(10^{-4}\, M_\odot)$ | $^{13}$C mass fraction |
| Three-zone | 9.38 | D4.5 to D1.5 |
| Three-zone_p2 | 18.76 | D6 to D1.5 |
| Zone-II_p2 | 10.6 | D6 to D1.5 |
| Zone-II_p4 | 21.2 | D7.5 to D1.5 |

Note: The constraints are for 0.5 $Z_\odot$ AGB stars.

Table 5. Effects of Nuclear Uncertainties on the Constraints in Table 4.

| Case 1: ½×K94 90% KAD $^{86}$Sr MACS | | Case 2: K94 110% KAD $^{86}$Sr MACS | |
|---|---|---|---|
| Models | $^{13}$C mass fraction | Models | $^{13}$C mass fraction |
| Three-zone | D3 to D1.3 | Three-zone | D4.5 to D1.5 |
| Three-zone_p2 | D4.5 to D1.3 | Three-zone_p2 | D6 to D2 |
| Zone-II_p2 | D3 to D1.5 | Zone-II_p2 | D4.5 to D1.5 |
| Zone-II_p4 | D6 to D1.5 | Zone-II_p4 | D7.5 to D1.5 |

Table 6. Further Constraints on Table 4 from $\delta(^{135}Ba/^{136}Ba)$ in Mainstream SiCs.

| Models | $^{13}$C-Pocket Mass ($10^{-4}\,M_\odot$) | Total $^{13}$C Mass ($10^{-7}\,M_\odot$) | | | | | | |
|---|---|---|---|---|---|---|---|---|
| | | D7.5 | D6 | D4.5 | D3 | D2 | D1.5 | D1.3 |
| Three-zone | 9.38 | | | | 6.38 | 9.57 | 12.76 | |
| Three-zone_p2 | 18.76 | | 6.38 | 8.50 | 12.76 | 19.14 | | |
| Zone-II_p2 | 10.6 | | | 7.53 | 11.30 | 16.96 | 22.61 | |
| Zone-II_p4 | 21.2 | | 11.28 | 15.08 | 22.61 | 33.92 | | |

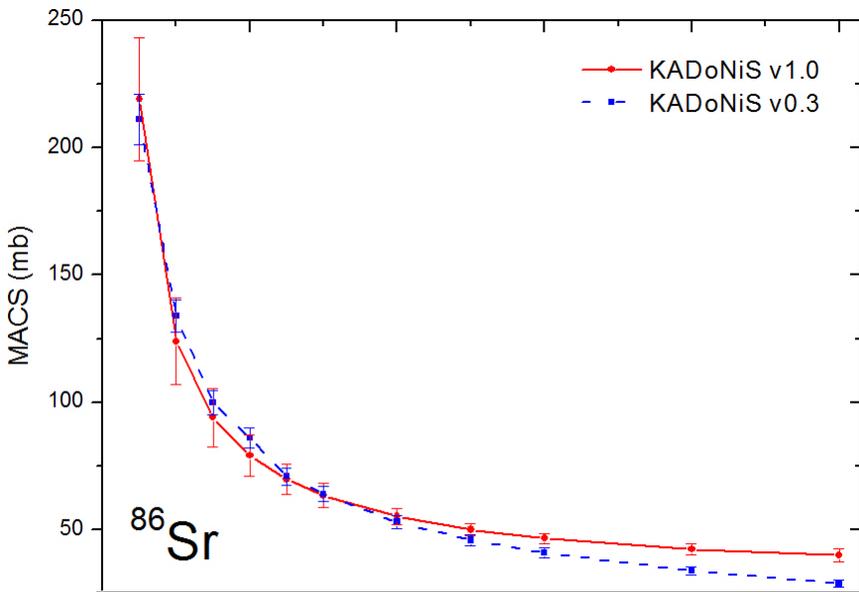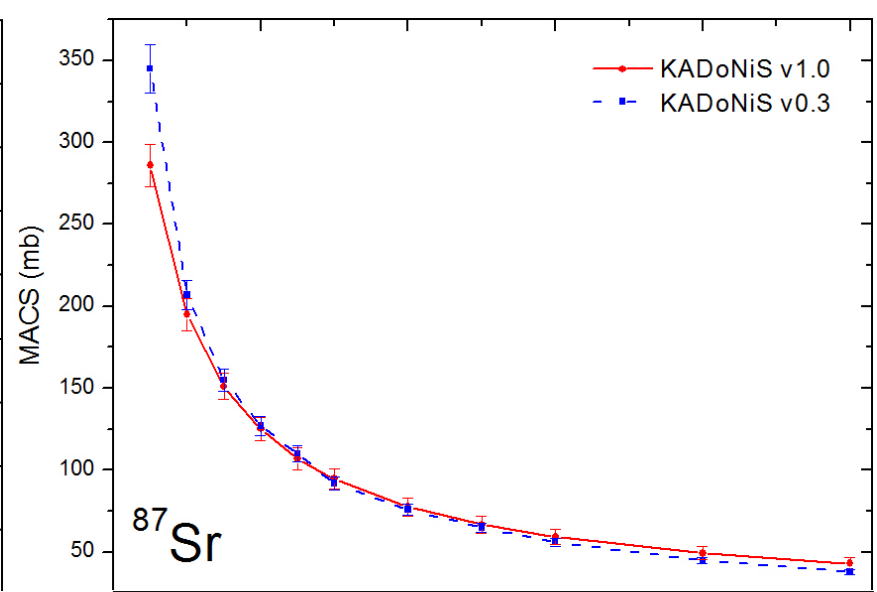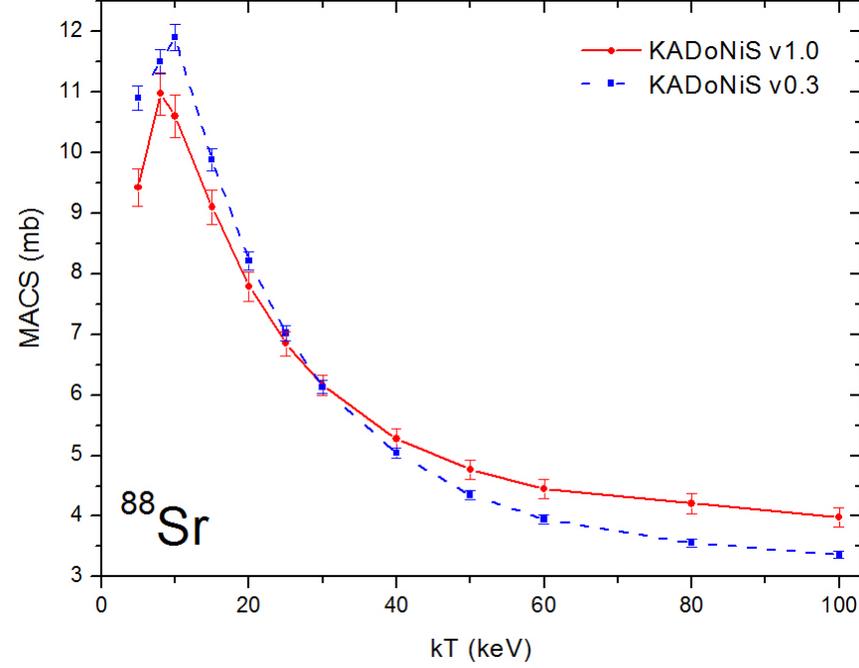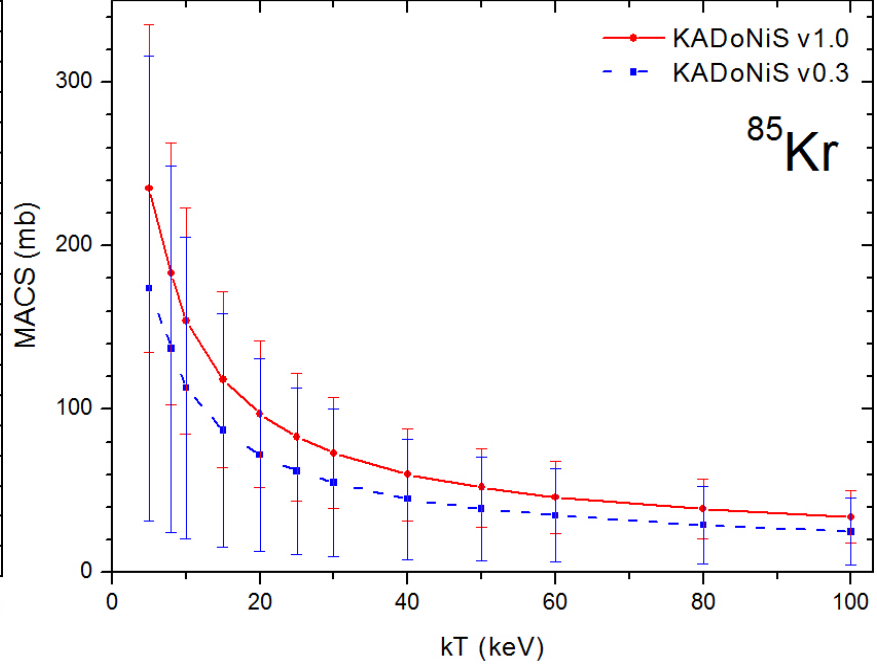

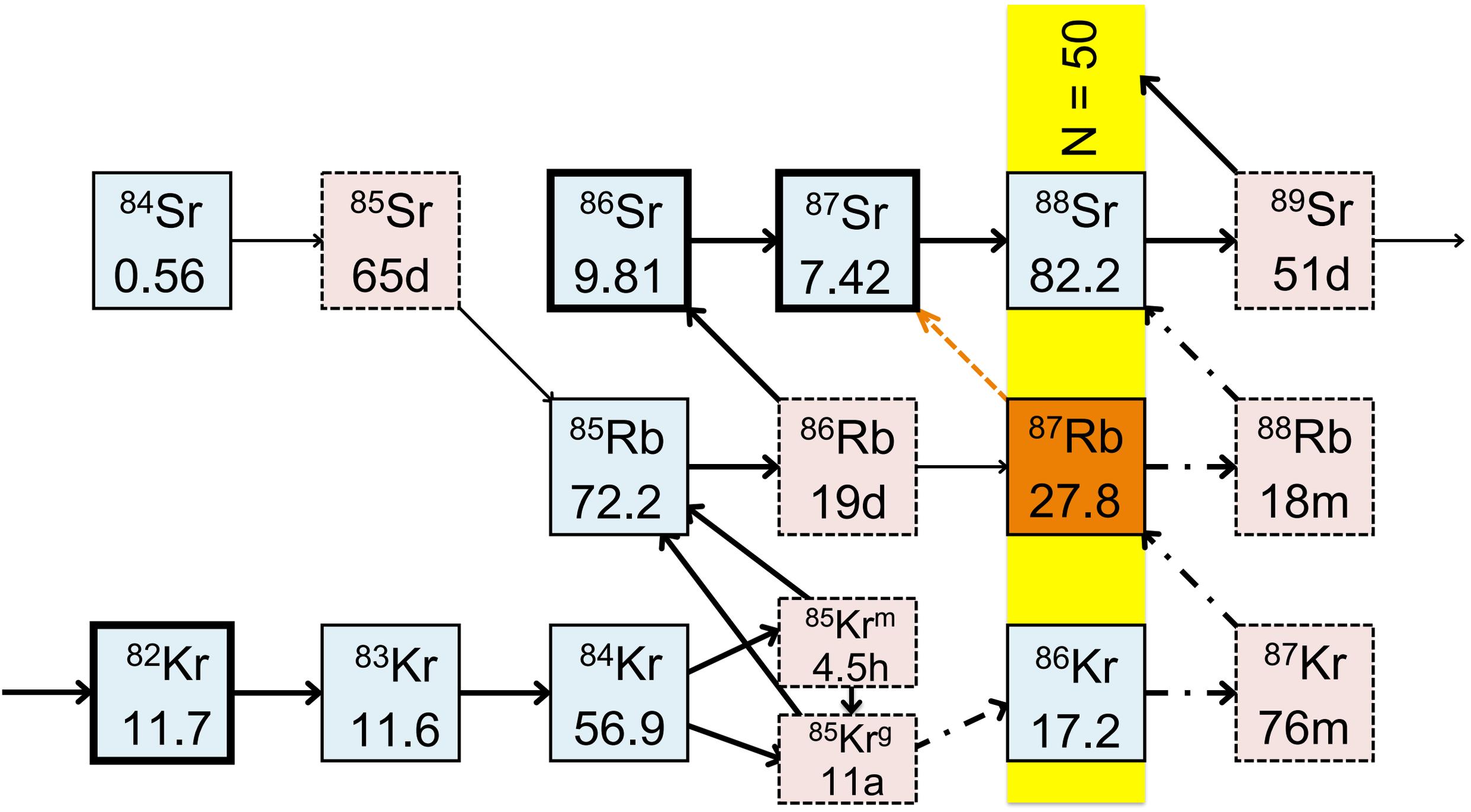

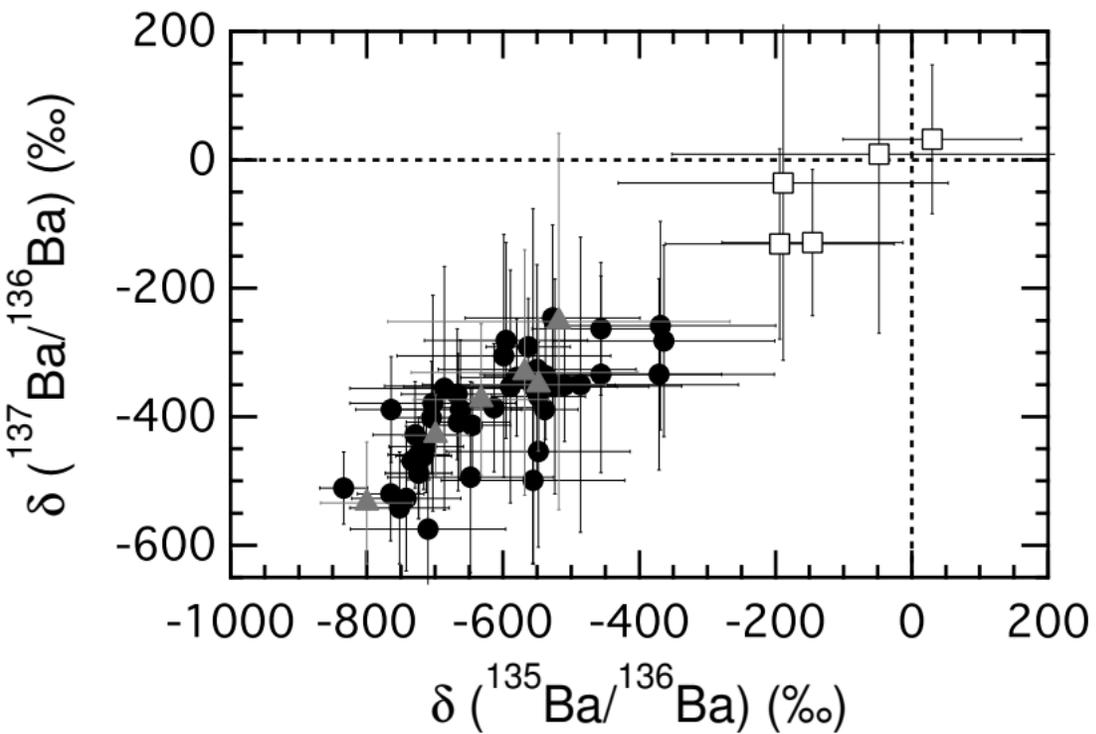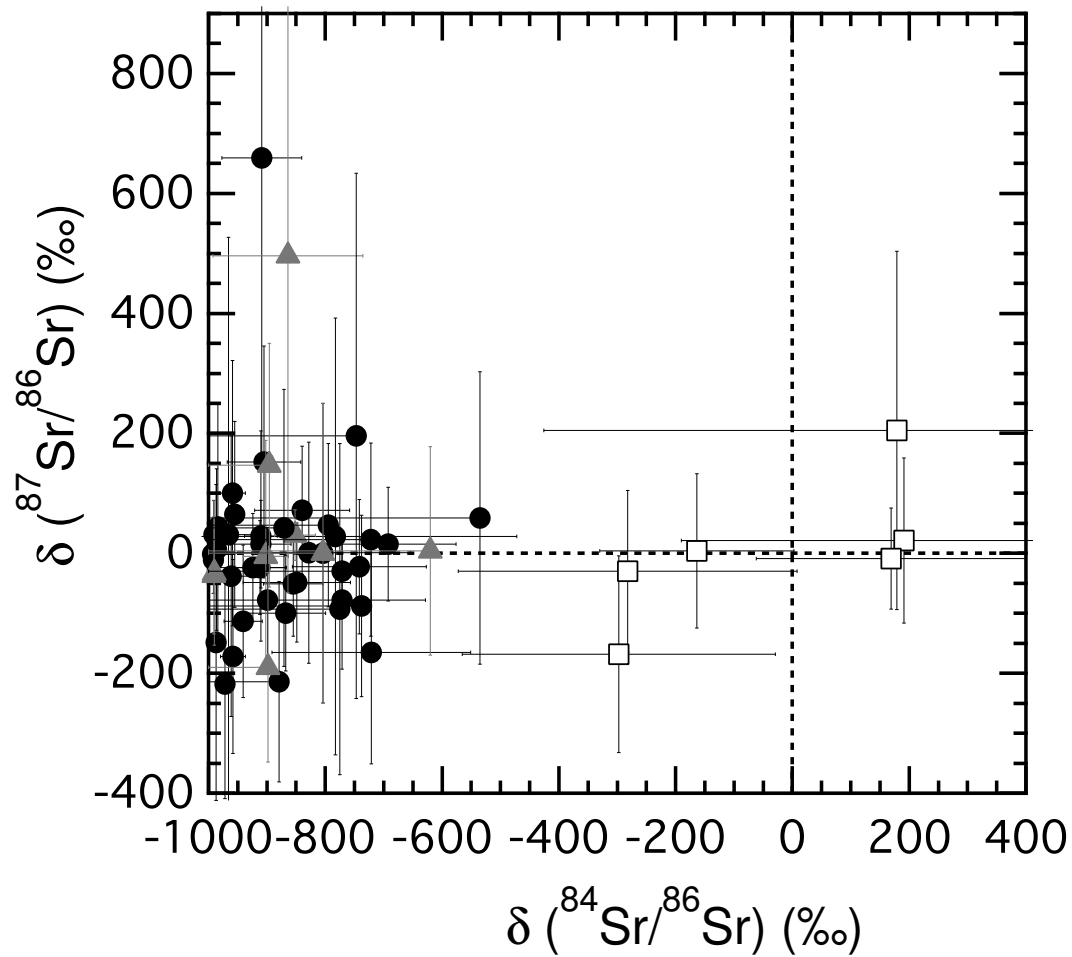

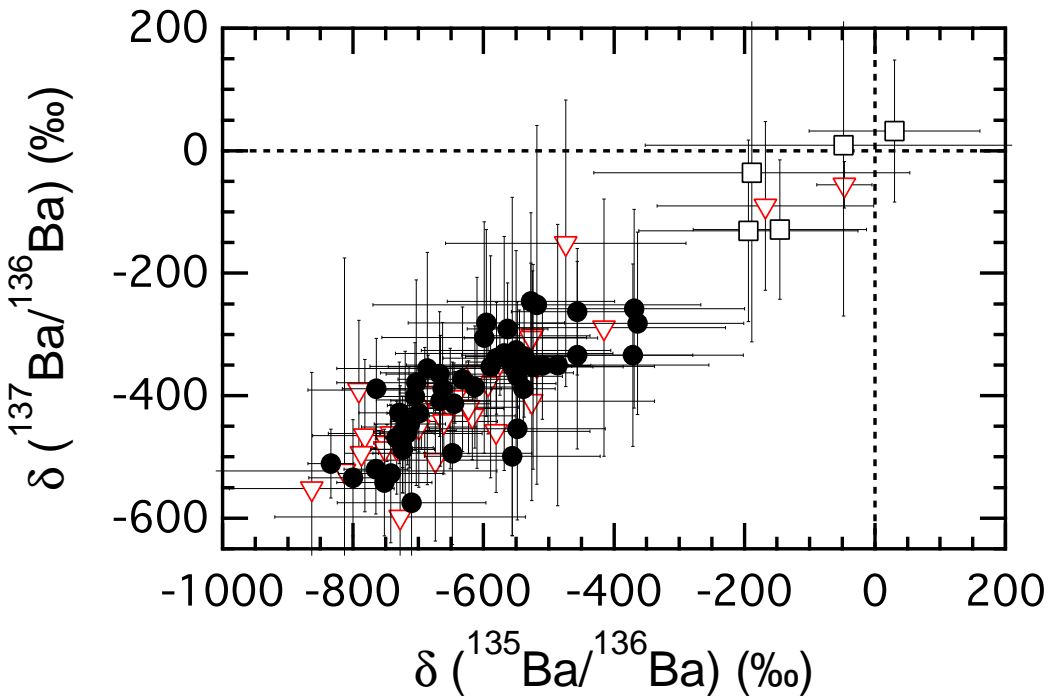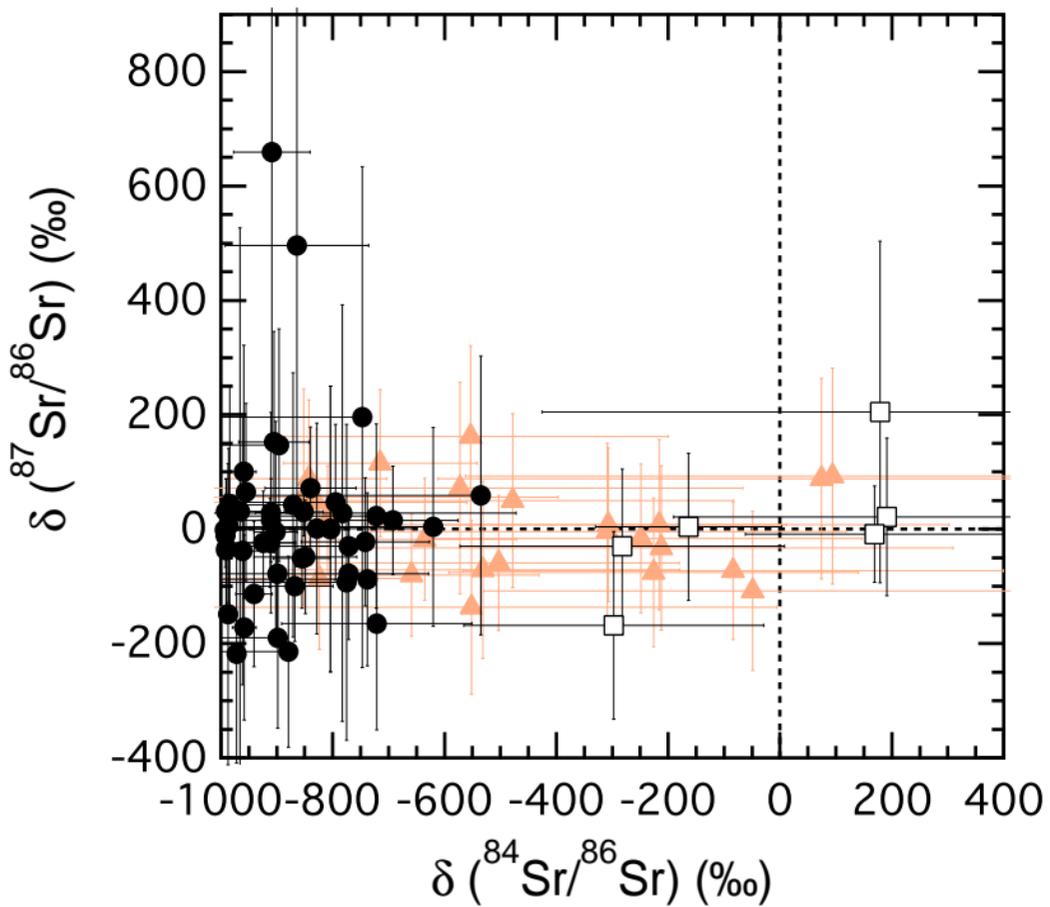

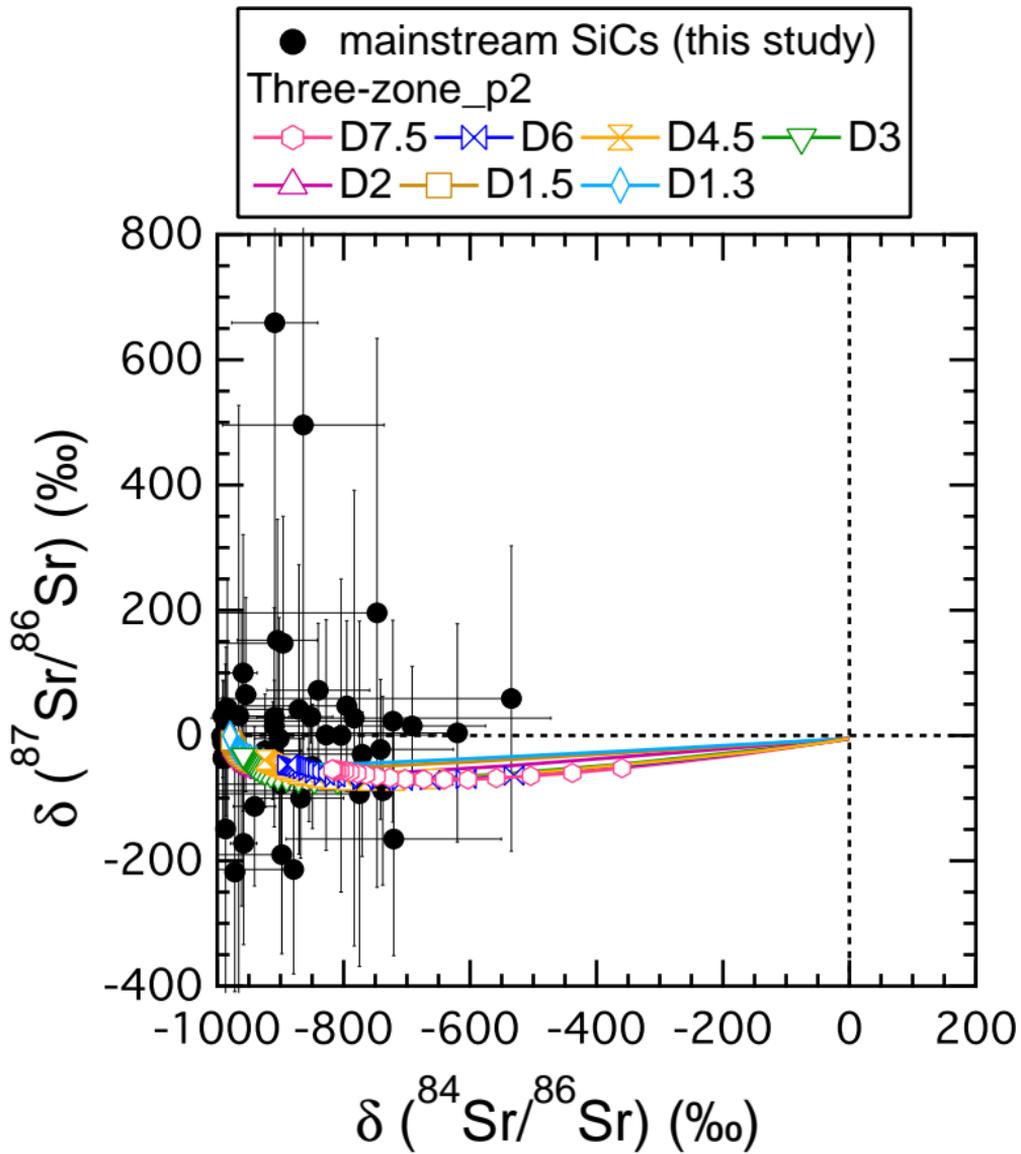

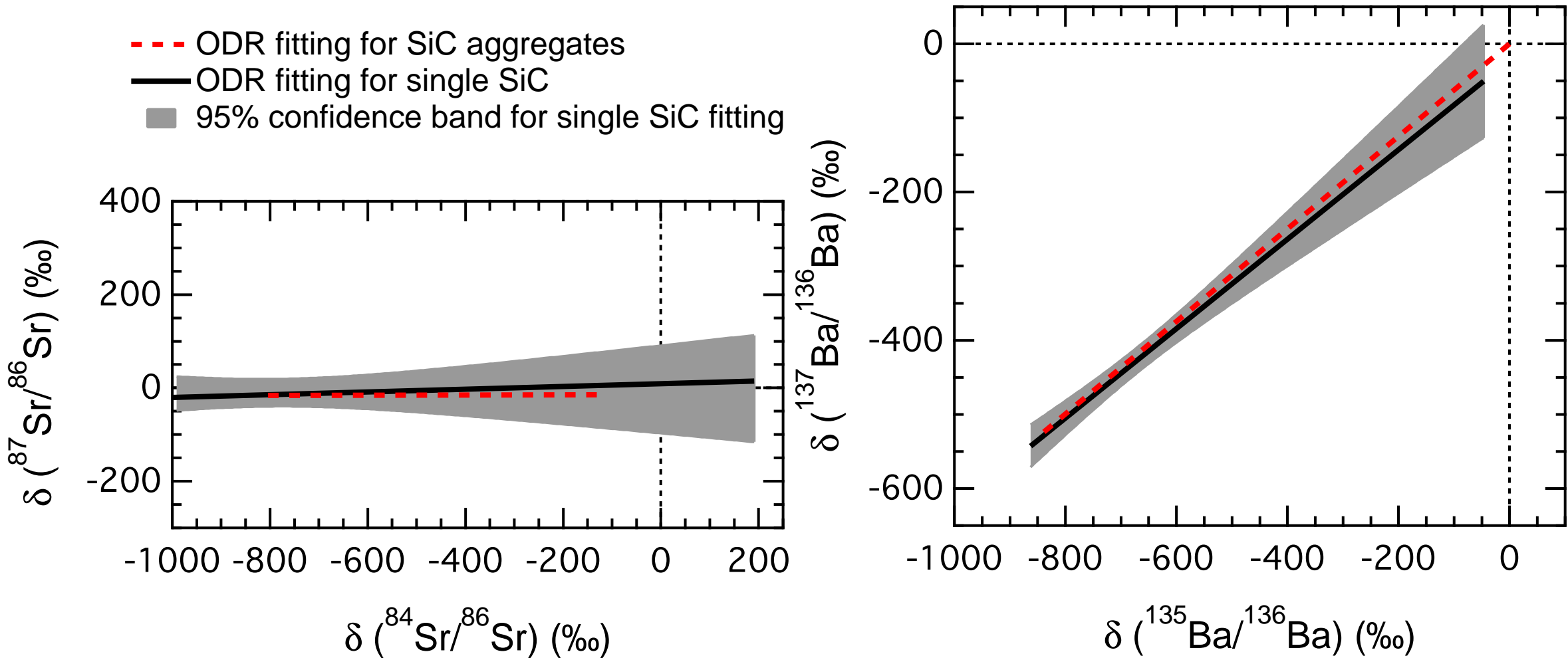

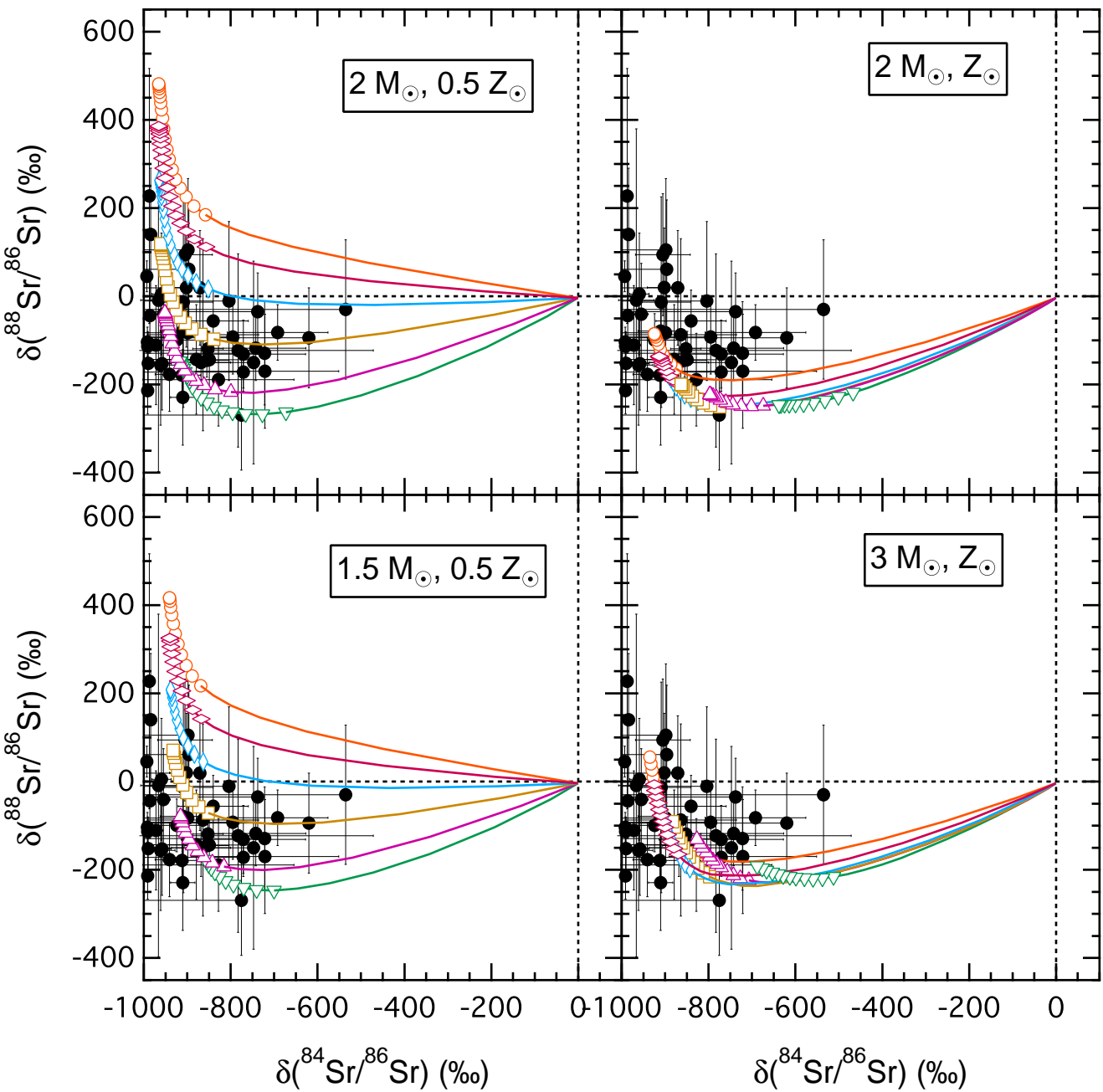

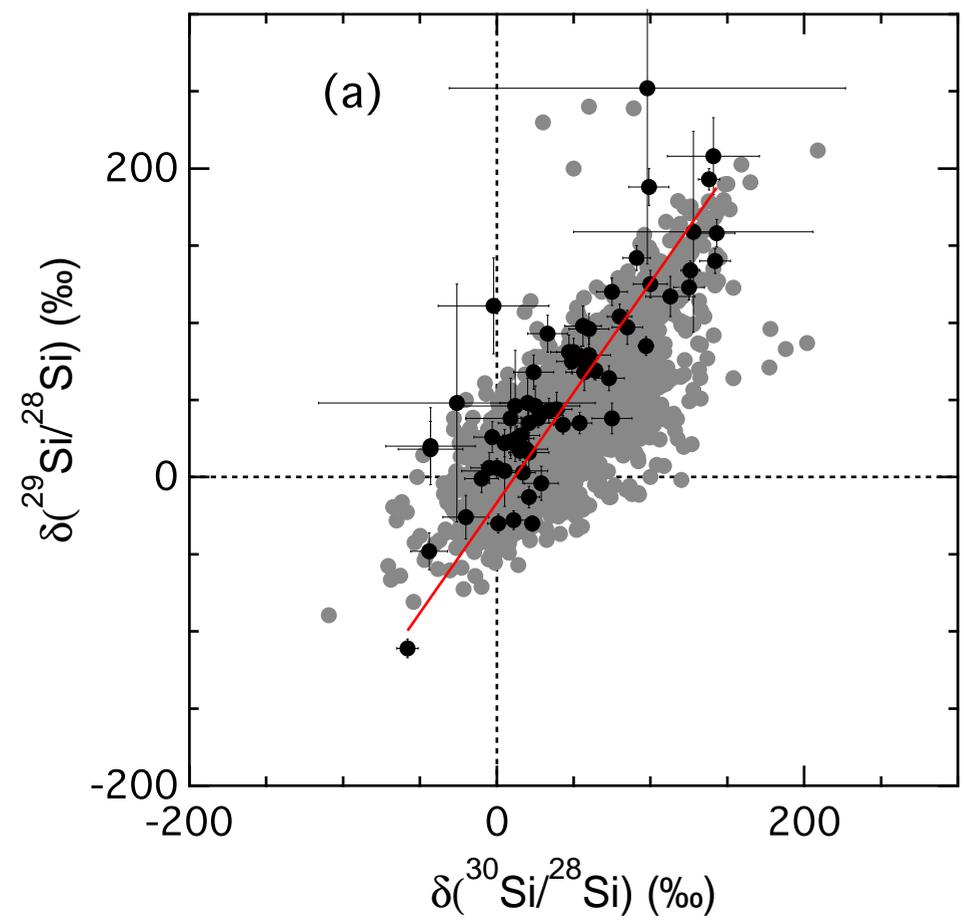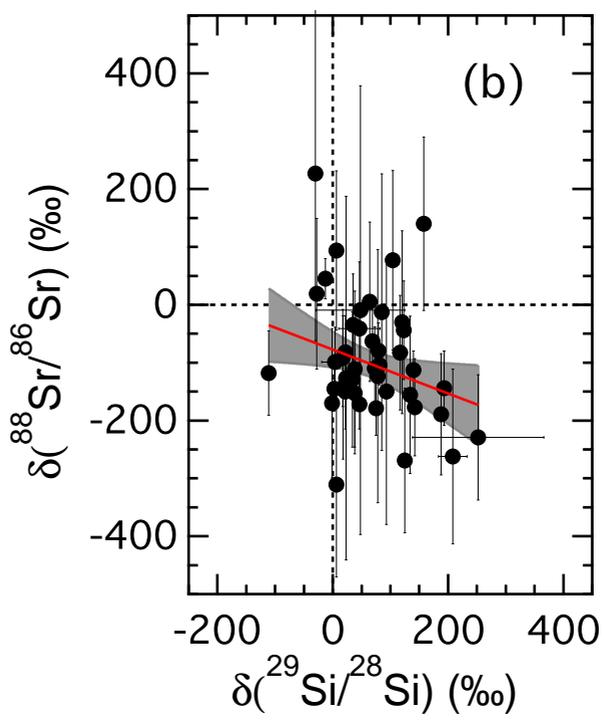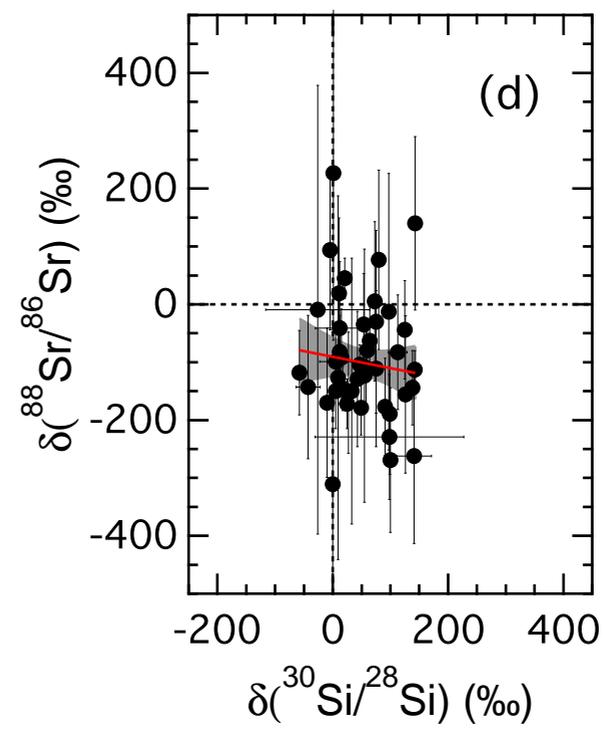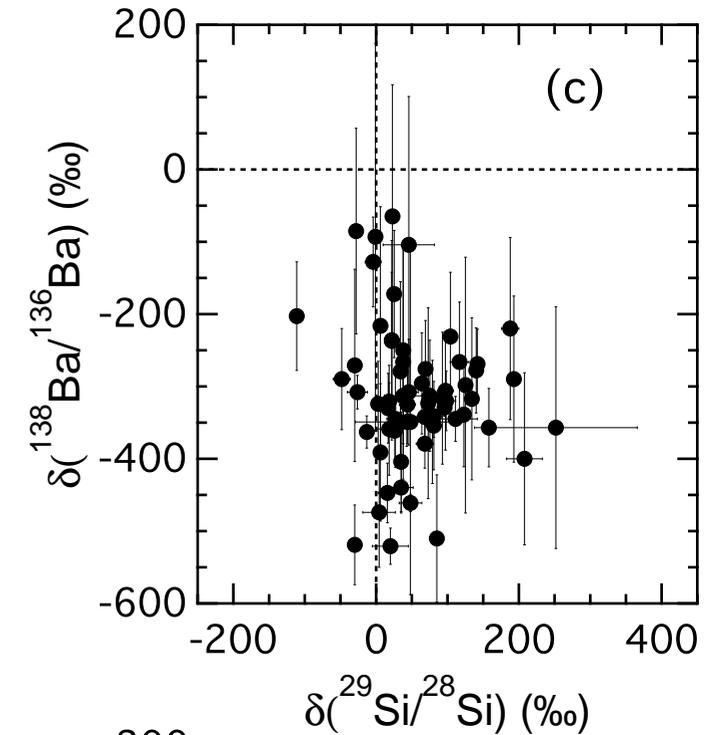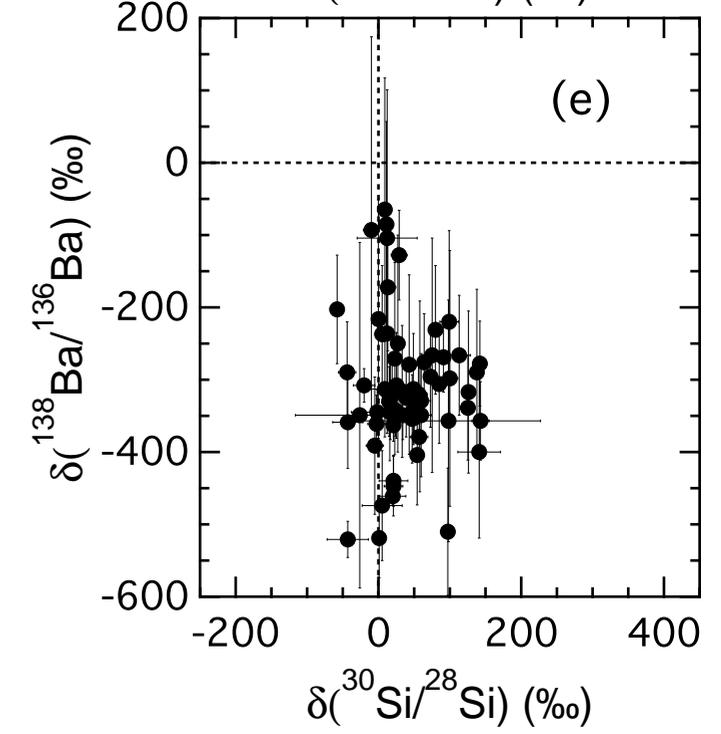

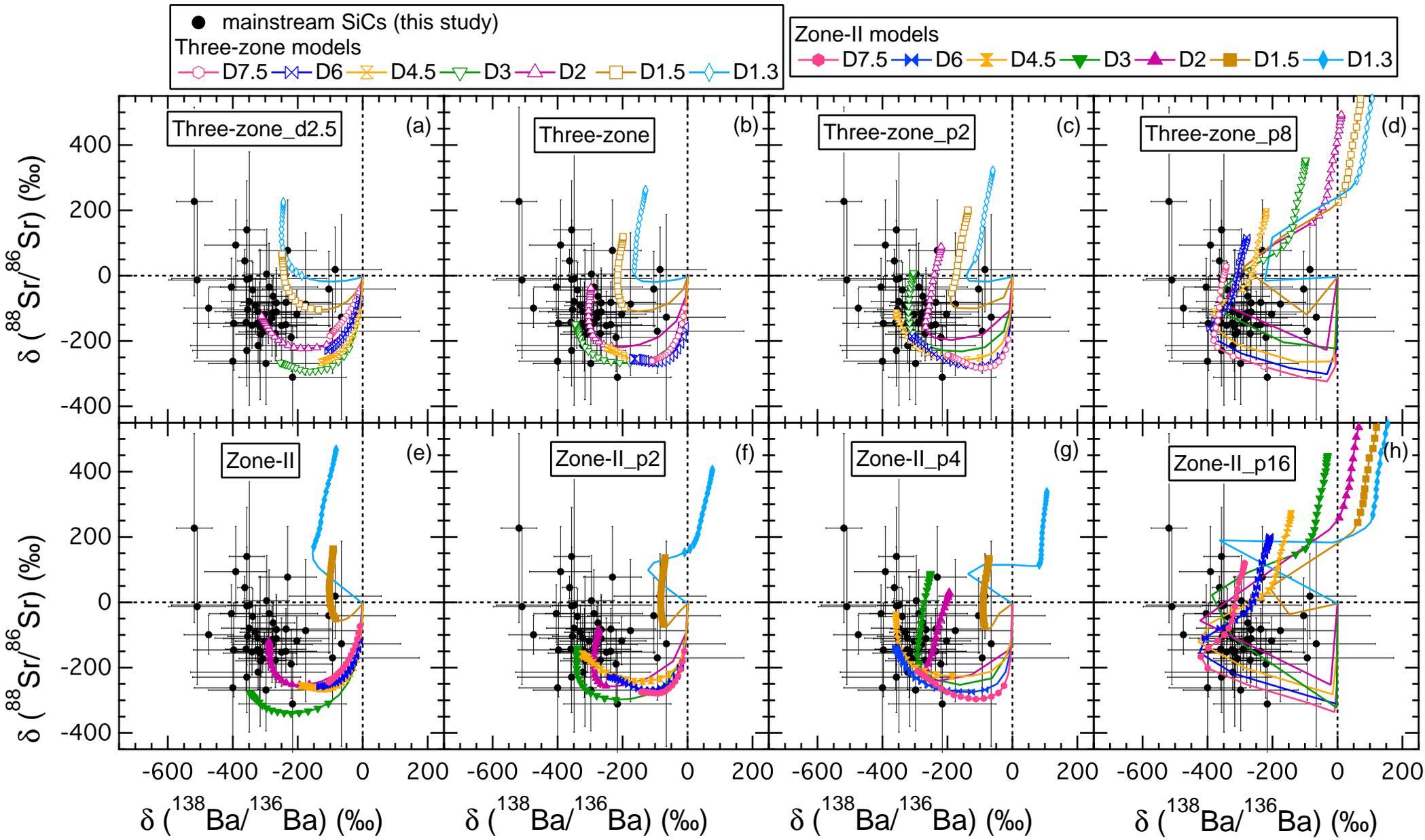

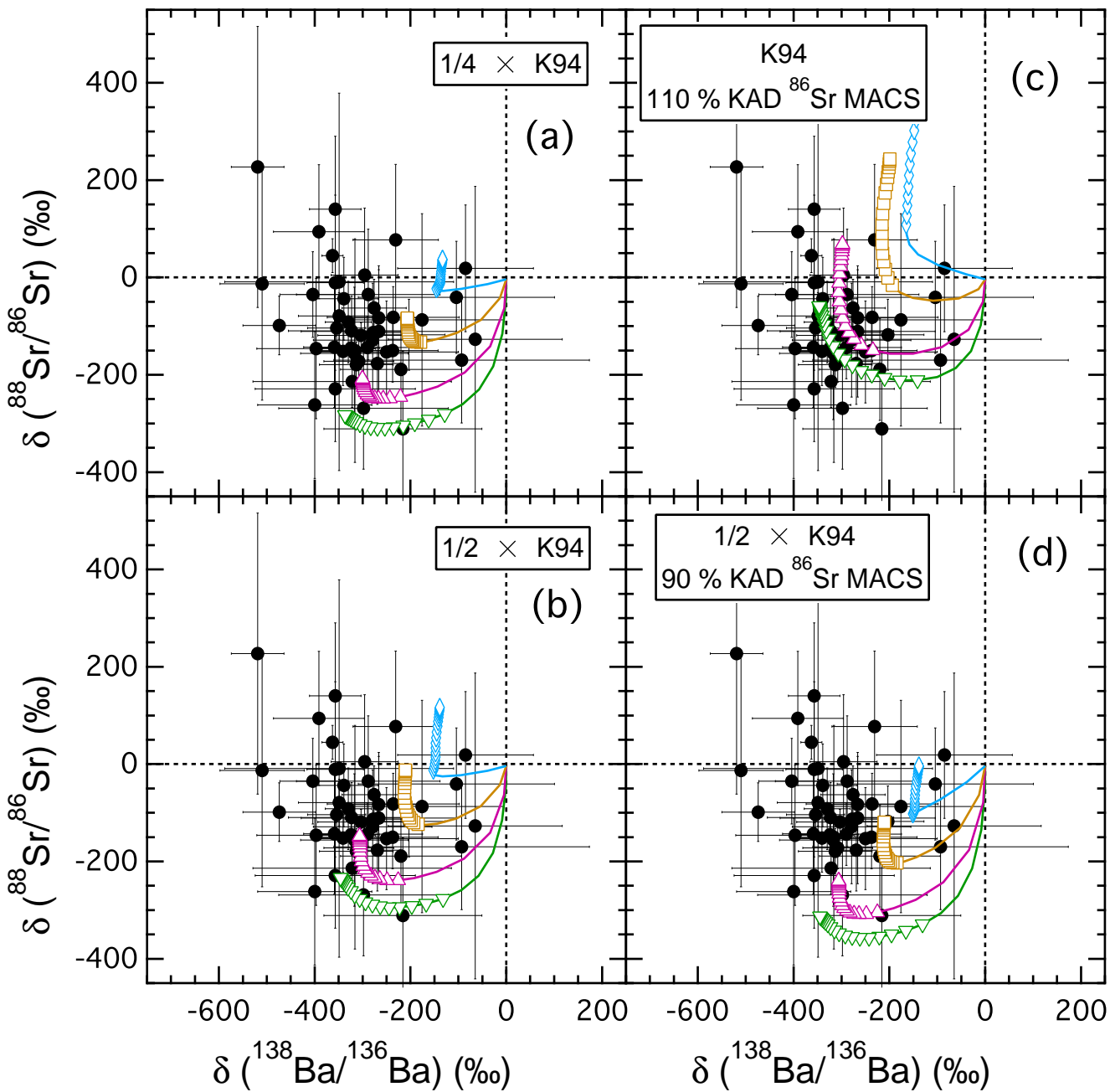

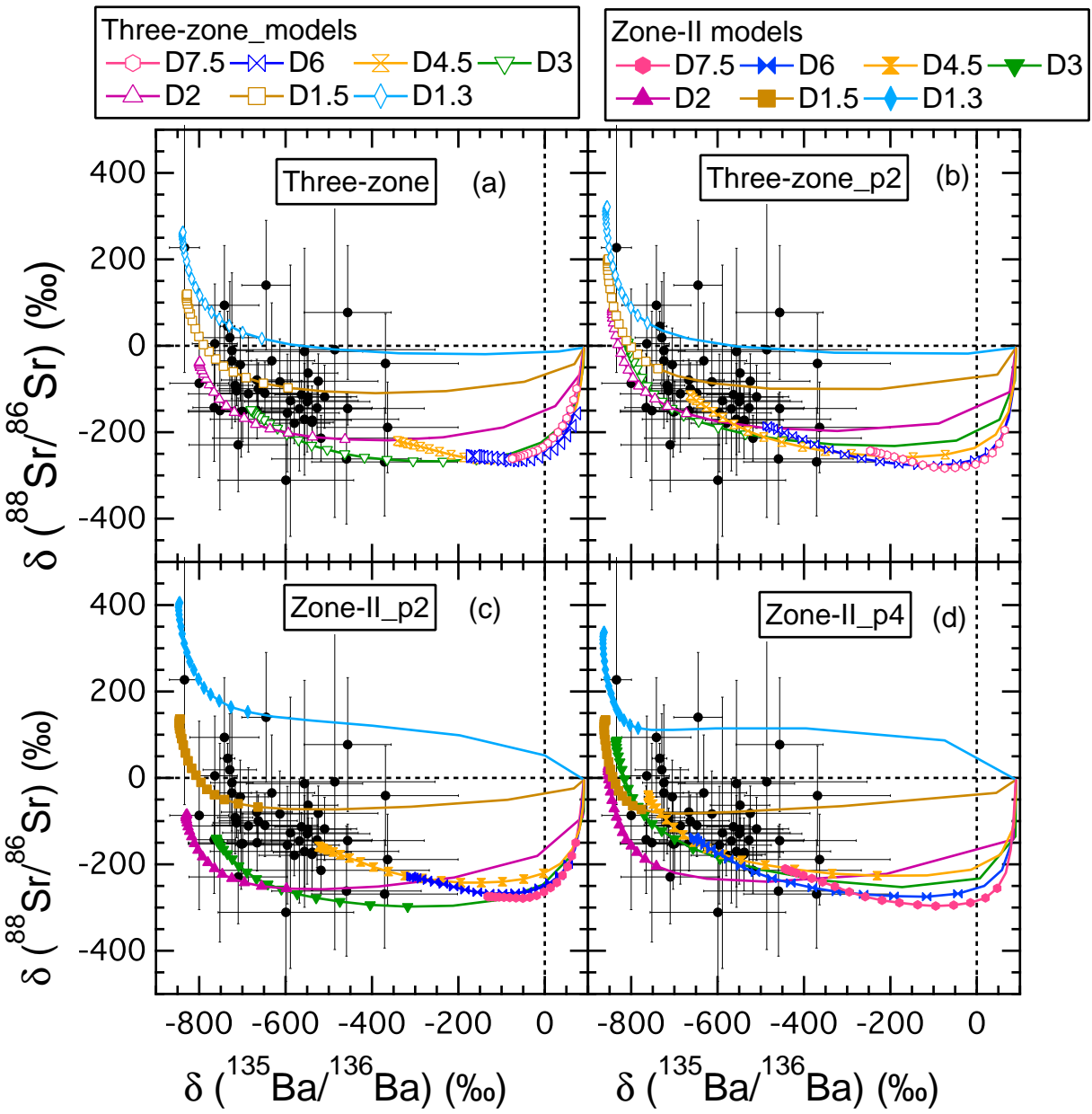

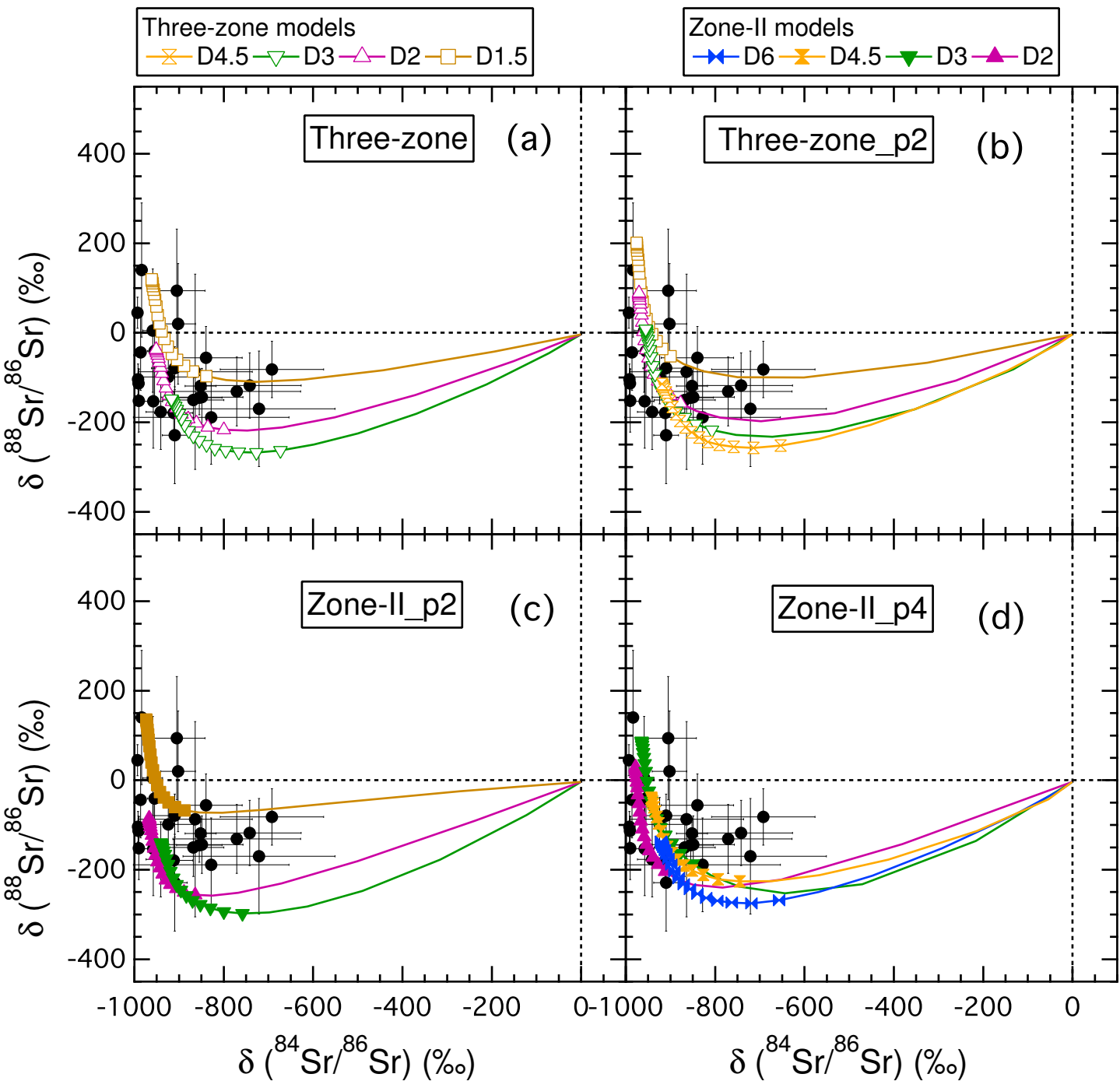

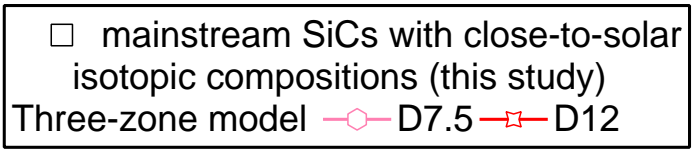
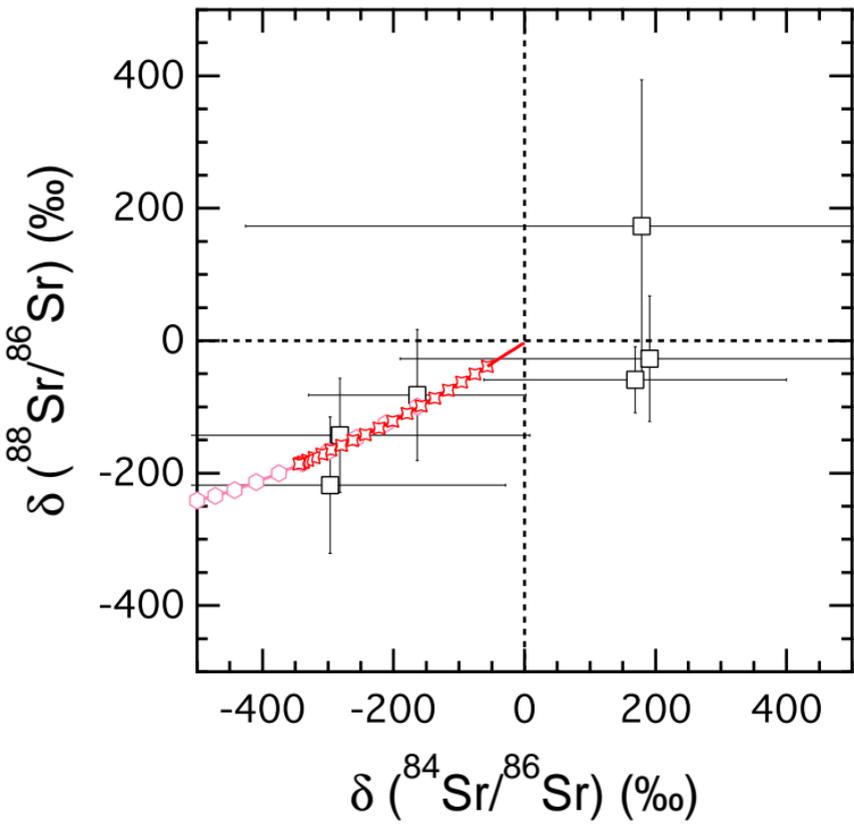
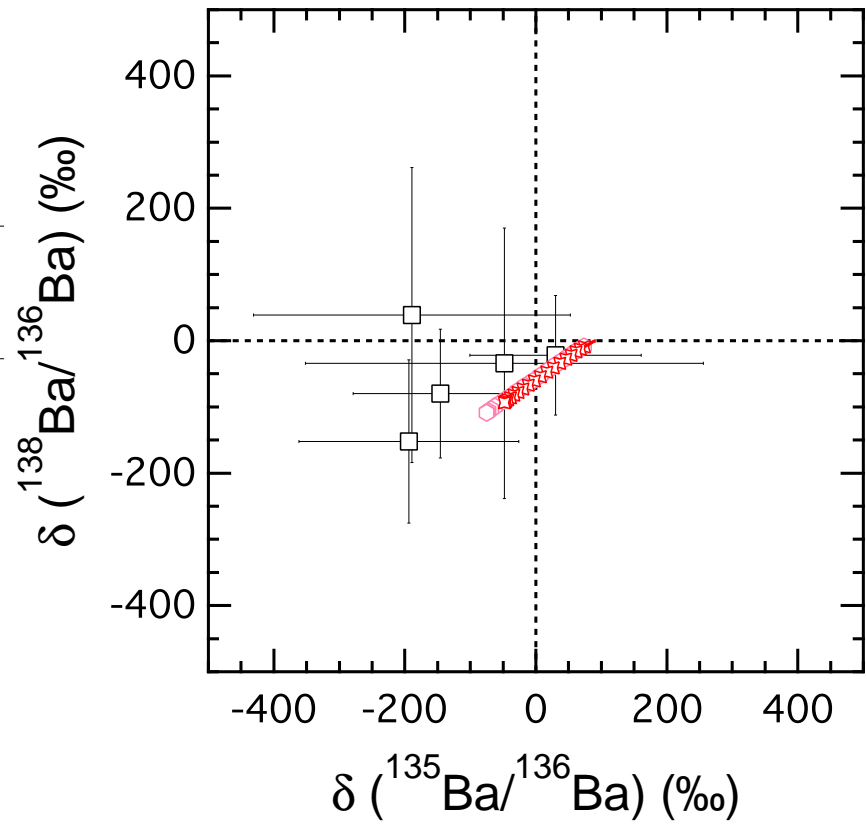

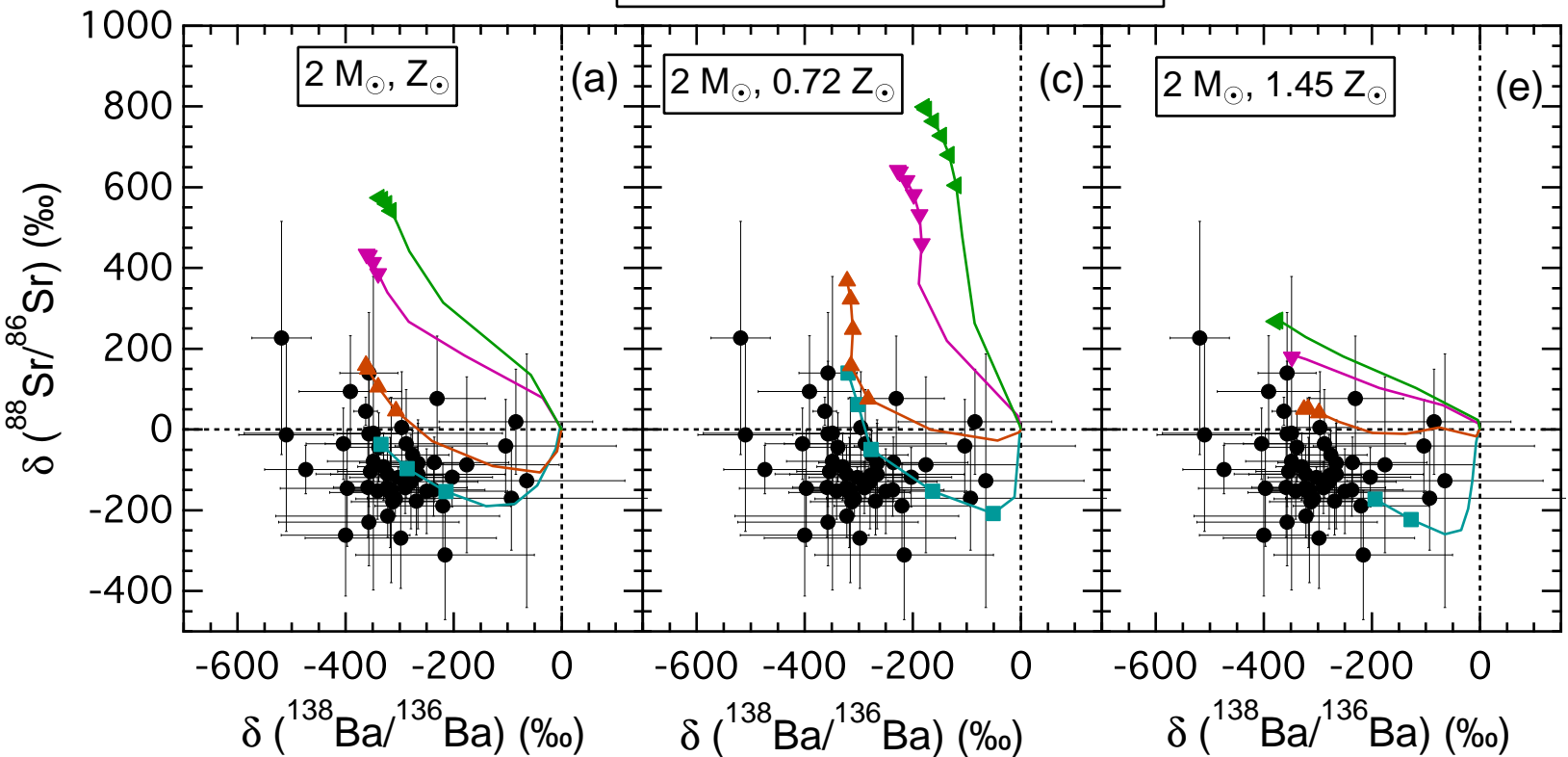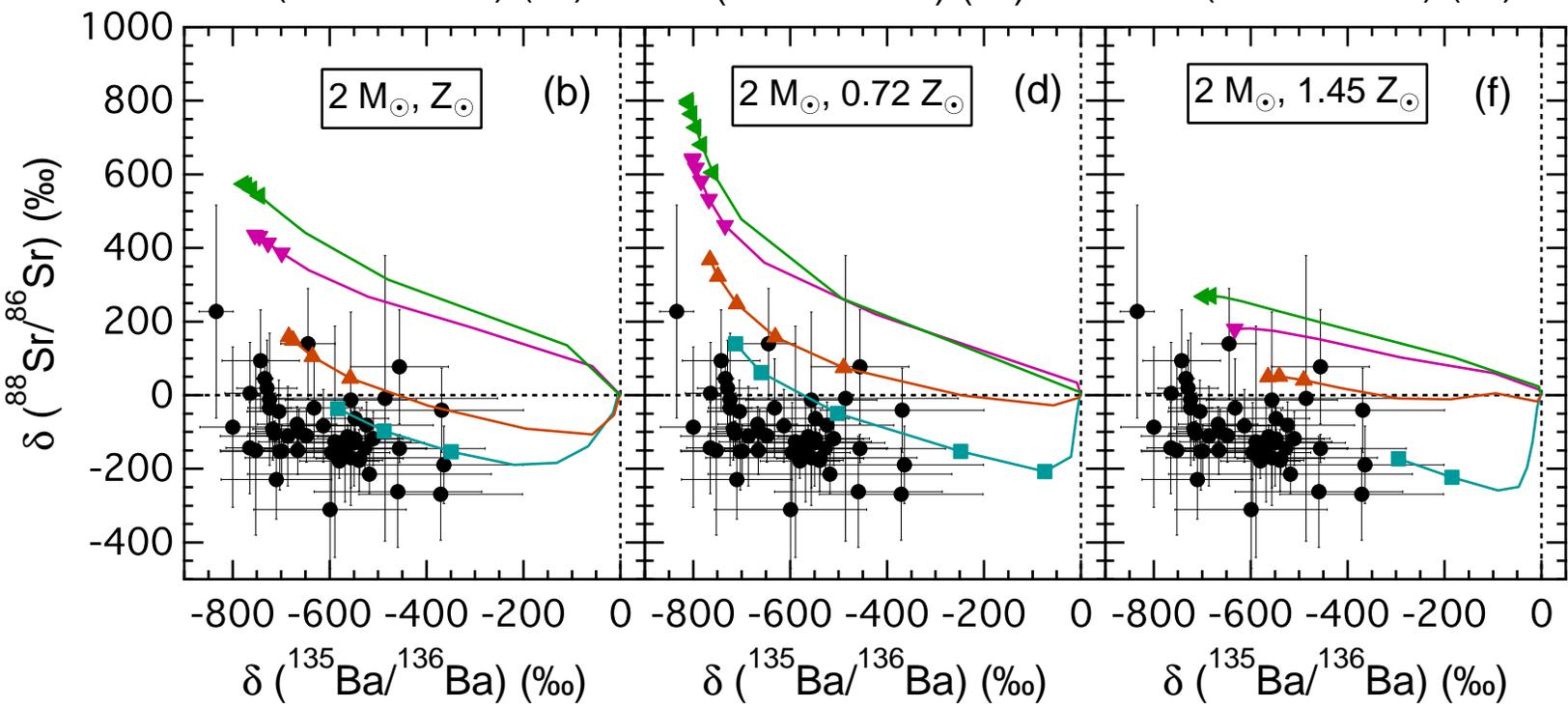

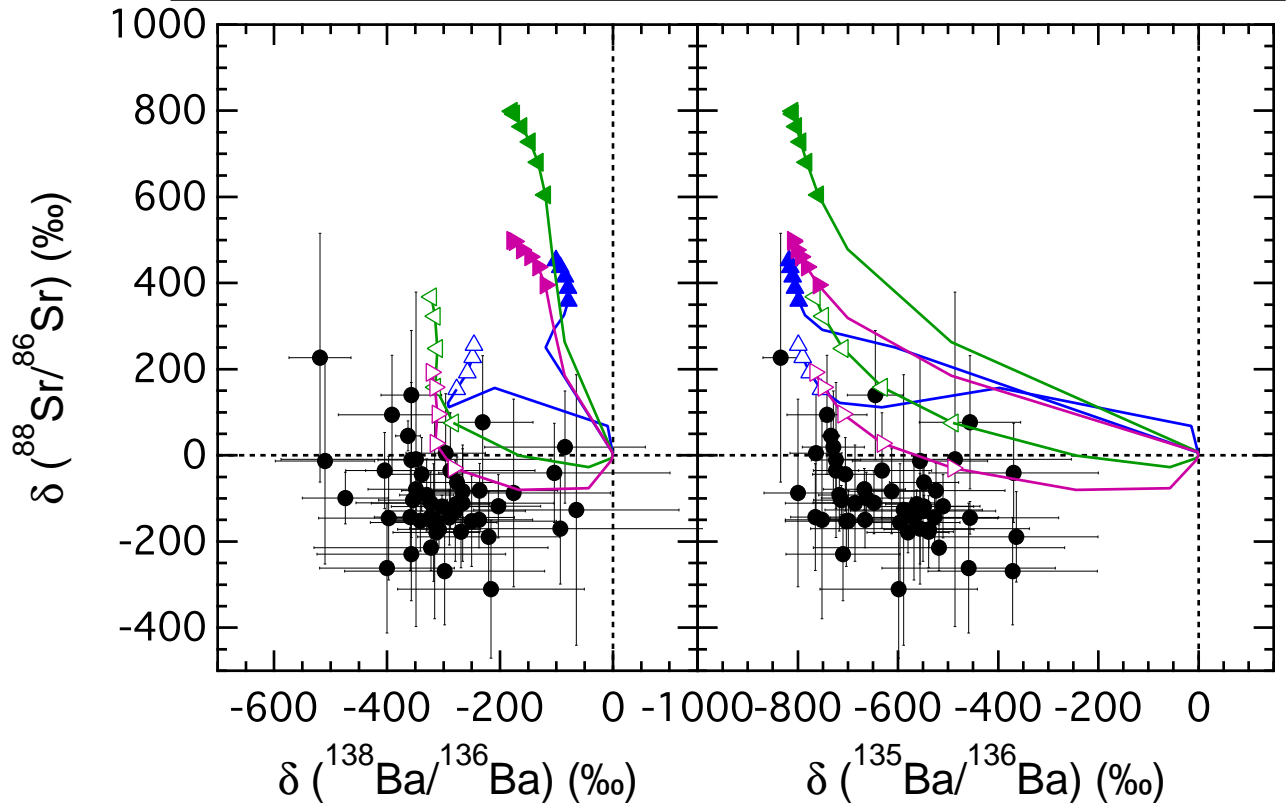